\documentclass[a4paper,11pt]{article}
\usepackage{graphicx} 
\pdfoutput=1 
\usepackage{jheppubMod} 
\usepackage{graphicx}
\usepackage{color}
\usepackage{array,arydshln}
\usepackage{amssymb,amsmath}
\usepackage{tabularray}
\usepackage{physics}
\usepackage{mathtools}
\usepackage[T1]{fontenc} 
\usepackage{float}

\newcommand{\GeV}{{\text{GeV}}}

\definecolor{navy}{rgb}{0.9,1.,1}

\def\bal#1\eal{\begin{align}#1\end{align}}

\preprint{TU-1246}

\title{Thermal Production of Axions from Heavy Higgs Bosons}

\author{
{\large Kodai Sakurai$^{1,2}$, Fuminobu Takahashi$^{1}$}
\\*[20pt]
{\it \normalsize 
$^1$ Department of Physics, Tohoku University, 
Sendai, Miyagi 980-8578, Japan} \\*[5pt]

{\it \normalsize 
$^2$ Institute of Theoretical Physics, Faculty of Physics, University of Warsaw, ul. Pasteura 5, PL-02-093 Warsaw, Poland} \\*[5pt]

}

\emailAdd{kodai.sakurai.e3@tohoku.ac.jp}
\emailAdd{fumi@tohoku.ac.jp}

\vspace{20pt}

\abstract{
We discuss the thermal production of axions in renormalizable models involving two Higgs doublet fields and a complex singlet field with a global $U(1)$ Peccei-Quinn symmetry, i.e., DFSZ type axion models. 
We demonstrate that, when the reheating temperature exceeds the mass scale of heavy Higgs bosons, axions are efficiently produced through heavy Higgs boson decays and scatterings at temperatures comparable to the heavy Higgs boson mass scale.
As a result, the abundance of thermally produced axions is independent of the reheating temperature, which should be contrasted with the KSVZ axion model.
This is because thermal productions via renormalizable interactions are IR-dominated processes. 
We demonstrate that the heavy Higgs boson decays are the main channels for axion thermal productions among various processes in 
the DFSZ-type axion models, which were missed in the literature.
Our results apply to the original DFSZ QCD axion model since the production mechanism does not depend on the axion mass.
As an application of axion productions from the heavy Higgs boson decays, 
we calculate the contributions to $\Delta N_{\rm eff}$ for axions with a mass smaller than ${\cal O}(0.1){\rm eV}$. {Future measurements of $\Delta N_{\rm eff}$ could constrain model parameters in both axion and Higgs sectors.}
Focusing on axions with masses from keV to sub-GeV scale, we then discuss how cosmological observations such as X-ray and cosmic microwave background constrain the produced axion. 
We show that a large portion of the parameter space of the models can be explored even if the amount of the axion produced from the heavy Higgs bosons is much smaller than the observed cold dark matter abundance. 
  }

\begin{document}

\maketitle

\newpage
\section{Introduction}
Axions, a type of Nambu-Goldstone (NG) boson, arise from the spontaneous breaking of a global symmetry and have attracted considerable attention in particle physics and cosmology. These hypothetical particles are considered prime candidates for dark matter in the universe. In particular, 
the QCD axion~\cite{Weinberg:1977ma,Wilczek:1977pj} is a NG boson of a U(1) Peccei-Quinn (PQ) symmetry~\cite{Peccei:1977hh,Peccei:1977ur}, and is the most plausible solution to the strong CP problem in the Standard Model (SM). See Refs.~\cite{Kim:2008hd,Arias:2012az,Marsh:2015xka,DiLuzio:2020wdo,OHare:2024nmr} for comprehensive reviews. More generally, string theory is known to predict many axions at low energies, which can have similar couplings to QCD axions but may exhibit a wider range of masses.

In this paper, we mainly focus on axions with masses ranging from keV to sub-GeV scales. This mass range is particularly intriguing because it offers a rich landscape of detection possibilities and cosmological implications.
For axions in the ${\cal O}(1-10^3)$\,keV mass range, direct detection experiments can probe the axion-electron coupling~\cite{XENON:2022ltv,COSINE-100:2023dir} by looking for absorption signals. Additionally, indirect searches using X-ray detectors, such as XRISM~\cite{XRISMScienceTeam:2020rvx}, can look for spectral lines resulting from axion decay into two photons (see also Refs.~\cite{Higaki:2014zua,Jaeckel:2014qea}). These line searches provide a complementary approach to direct detection methods and can probe a significant portion of the axion parameter space.
As we move towards higher masses, cosmological probes become increasingly important. Axions in the sub-MeV to sub-GeV range can have significant impacts on cosmic microwave background (CMB) anisotropies and Big Bang Nucleosynthesis (BBN).  Furthermore,  axions can be copiously produced in the hot interior of stars, potentially influencing stellar evolution~\cite{Giannotti:2017hny}. These observations provide powerful constraints on axion properties.

For axions to explain dark matter, they must be produced in sufficient quantities in the early universe. One of the most well-established production mechanisms is the misalignment mechanism~\cite{Preskill:1982cy,Abbott:1982af,Dine:1982ah}. This process is inherent to axion cosmology and occurs since the axion field is naturally displaced from its minimum in the early universe. As the Hubble parameter decreases, the axion begins to oscillate around the minimum, effectively producing a condensate of cold axions.\footnote{Several variants  have been proposed, such as the kinetic misalignment~\cite{Co:2019jts}, the trapped misalignment~\cite{Higaki:2016yqk,Nakagawa:2020zjr,DiLuzio:2021gos,Jeong:2022kdr}, and the bubble misalignment mechanism~\cite{Lee:2024oaz,Nakagawa:2022wwm}. } Interestingly, the axion with a mass in the ${\cal O}(1-10)$ keV range as well as a decay constant of ${\cal O}(10^{10})$ GeV has an advantage that the right abundance can be naturally produced by the misalignment mechanism, and it can be searched for by the direct dark matter search experiment, the X-ray line search, and stellar cooling observations~\cite{Nakayama:2014cza,Takahashi:2020bpq}.

Another crucial production channel is thermal production from the hot plasma of SM particles in the early universe. While thermal production of axions has been studied extensively (e.g.~\cite{Masso:2002np,Graf:2010tv,Salvio:2013iaa,Arias-Aragon:2020shv}), these investigations often implicitly assume that axion couplings to SM particles arise in a KSVZ-like setup. In the KSVZ model~\cite{Kim:1979if, Shifman:1979if}, axions couple to heavy PQ quarks, which can be integrated out at low energies, resulting in effective axion couplings.
Although the KSVZ axion model is simpler than the DFSZ model~\cite{Dine:1981rt,Zhitnitsky:1980tq}, which requires two or more Higgs doublets, the difference between these models extends beyond mere complexity. In fact, the thermal production of axions differs significantly between the KSVZ and DFSZ models. This distinction is crucial for accurately predicting the cosmological abundance of axions and their subsequent phenomenology.

The primary goal of this paper is to thoroughly investigate the production of axions from thermal plasma in the framework of  DFSZ-like axion models. We will carefully study how axions are produced from interactions with extra Higgs particles, a feature unique to the DFSZ model. By doing so, we aim to provide a more comprehensive understanding of axion production mechanisms and their cosmological implications. 

Let us emphasize that the thermal production of axions from the thermal bath in the DFSZ axion model we examine is highly general and applicable regardless of whether the axion is a QCD axion or a general axion. Our analysis will include detailed calculations of the axion abundance and an exploration of various cosmological constraints that can be applied to this scenario. Particularly interesting are the cosmological constraints, which lead to restrictions in the keV to sub-GeV mass range, crucial for direct and indirect detection. However, we emphasize that the production processes and axion abundance calculations themselves are applicable to a much broader range of masses.

{As a related work of this paper, thermal productions of axions in the DFSZ model have been studied under the assumption that axions are thermalized in Ref.~\cite{DEramo:2021lgb}. 
In this reference, the threshold effects of QCD, electroweak, and heavy Higgs bosons are discussed by computing not only the axion productions from SM particle scatterings but also heavy Higgs boson scatterings (see also relevant works considering hadron scatterings~\cite{Chang:1993gm,Hannestad:2005df,DEramo:2014urw,Kawasaki:2015ofa,Ferreira:2020bpb,Giare:2020vzo,DiLuzio:2021vjd,Notari:2022ffe}, lepton scatterings~\cite{DEramo:2018vss,Badziak:2024szg,Badziak:2024qjg} ). 
In contrast to previous works, we consider freeze-in productions~\cite{Hall:2009bx} of axions from heavy Higgs bosons, including their decays.
As we will see later, contributions from heavy Higgs boson decays are almost ten times larger than those from the scatterings of heavy Higgs bosons. 
Although previous works mostly consider ultralight axions such that they contribute to the effective number of neutrino species, we also discuss the cosmological implications of the thermal axion productions to the keV to suh-GeV mass range. 
}

Lastly, let us comment on a related work conducted in the context of supersymmetric extensions of the SM. In supersymmetric theories, the axionic superpartner, known as the axino, has a mass that depends on how supersymmetry breaking is mediated to the SM sector. Previous studies~\cite{Chun:2011zd,Bae:2011jb,Choi:2011yf,Bae:2011iw} have investigated the production of axinos from the decay and scattering of Higgs particles in the supersymmetric DFSZ axion model. While our focus is on axions rather than axinos, some of the production processes we examine are analogous to those explored in these earlier works.

The rest of this paper is as follows. 
{In Sec.~\ref{sec:model}, we describe the renormalizable model that we focus on 
Sec.~\ref{sec:axion productions} is devoted to the formulations of the freeze-in productions of axions from the heavy Higgs boson decays and scatterings. 
The applications of axion productions from heavy Higgs bosons follow. 
In Sec.~\ref{sec:delta_Neff}, we discuss the contributions to $\Delta N_{\rm eff}$ and 
in Sec.~\ref{sec:cosmological constraints}, cosmological bounds for the axion in keV to sub-GeV scale mass range. 
Conclusions are given in Sec.~\ref{sec:conclusion}. 
}

\section{Model (2HDM + complex singlet scalar)} \label{sec:model}
Here, we describe a benchmark model for the discussion of axion thermal production from heavy Higgs bosons.
As a minimal setup involving both additional Higgs bosons and axions, we consider a model of singlet extension of two Higgs doublet models (2HDMs) with global $U(1)$ PQ symmetry.
Two Higgs doublet fields $\phi_{1,2}$ and a complex singlet field $S$ are charged under the $U(1)$ symmetry.
They transform under the $U(1)$ symmetry as
\begin{align}
\phi_{1,2}\to e^{iX_{1,2}\xi}\phi_{1,2},\quad
S\to e^{iX_{S}\xi}S,
\end{align}
We assume that the U(1) charge is conserved when
\begin{align}
{-X_{1}+X_{2}+2X_{S}=0\;.}
\end{align}
This allows us to include the terms ($S^2\phi^\dagger_1 \phi_2$+h.c.) in the Higgs potential, as we will shortly see. 
Similar to the DFSZ model~\cite{Zhitnitsky:1980tq,Dine:1981rt}, we assign the $U(1)$ charges for the Higgs fields and the complex singlet as
 \begin{align}
   X_1=2\sin^2\beta\;,\quad X_2=-2\cos^2\beta\;,\quad X_S=+1. 
  \end{align}
We assume that the $U(1)$ symmetry is only an approximate symmetry, which is explicitly broken by interactions outside the Higgs sector. As a result, the corresponding NG boson, an axion, acquires a small mass. We will discuss the origin of the axion mass later in this section.
In the following, we present the Higgs potential and Yukawa interactions for this setup. 

\subsection{The Higgs potential} 

The Higgs potential can be divided into three parts:
\begin{align}
V=V_{\phi}+V_{I}+V_{S}.
\label{eq:V_Z2}
\end{align}
The first term, $V_{\phi}$, describes the potential constructed from the Higgs doublet fields, while the third term, $V_{S}$, describes that of the complex singlet scalar field.
The second term, $V_{I}$, denotes the portal interactions between the Higgs doublet sector and the complex singlet sector.
Each part is given by
  \begin{align}
    V_{\phi} &= m_{11}^{\prime 2}(\phi_1^\dagger \phi_1) +  m_{22}^{\prime 2}(\phi_2^\dagger
    \phi_2)   \nonumber \\
     &+\lambda'_1(\phi_1^\dagger
    \phi_1)^2 + \lambda'_2(\phi_2^\dagger \phi_2)^2 
    + \lambda'_3 (\phi_1^\dagger \phi_1)( \phi_2^\dagger \phi_2) 
   + \lambda'_4 (\phi_1^\dagger \phi_2)( \phi_2^\dagger \phi_1) \;, \\
\label{e:potential}      
  V_{S}&=
  m^2_{S\bar{S}}|S|^{2}+\kappa_{}|S_{}|^{4} \; , \\
  \label{eq:V_I}
  V_{I}&=
  \kappa_{\phi 1}|S_{}|^{2}(\phi_{1}^{\dagger}\phi_{1})+\kappa_{\phi 2}|S_{}|^{2}(\phi_{2}^{\dagger}\phi_{2})
  +\kappa_{S}\big\{S^{2}(\phi_{1}^{\dagger}\phi_{2}^{})+{\rm h.c.}\big\} \; .
  \end{align}
  Although $\kappa_{S}$ is generally a complex parameter, it can be taken to be real by rephasing $\phi_2$ or $S$.
The component fields of these fields are written as
    \begin{eqnarray} 
      \phi_k = \frac{1}{\sqrt{2}}
      \begin{pmatrix} \sqrt{2}\, w_k^+ \\  v_k + h_k
        + i z_k  \end{pmatrix} \; (k = 1,2)\;,\quad
      S=\frac{1}{\sqrt{2}}(v_S+\rho)\exp\left(i\frac{\tilde{a}}{v_S}\right)\;.
      \label{eq:scalvev}
    \end{eqnarray}
  The vacuum expectation values (VEVs) $v_1$ and $v_2$ relate to the EW VEV and the mixing parameter for the Higgs doublet fields as $v^2_1+v^2_2=v^2=(246{\rm GeV})^2$ and $\tan\beta=v_2/v_1$. 

The typical mass scale of the radial mode of the singlet scalar $\rho$ is around $v_S$, which is taken to be sufficiently larger than the mass of heavy Higgs bosons in the following analysis. This allows $\rho$ to be integrated out.
The potential can be reduced after integrating out the CP-even component $\rho$ as
  \begin{align}
     V &= m_{11}^{ 2}(\phi_1^\dagger \phi_1) +  m_{22}^{ 2}(\phi_2^\dagger    \phi_2)
      -(m_{12}^2\phi_1^\dagger \phi_2+{\rm h.c.}) \nonumber \\
      &+\lambda_1(\phi_1^\dagger
     \phi_1)^2 + \lambda_2(\phi_2^\dagger \phi_2)^2 
     + \lambda_3 (\phi_1^\dagger \phi_1)( \phi_2^\dagger \phi_2) 
    + \lambda_4 (\phi_1^\dagger \phi_2)( \phi_2^\dagger \phi_1)  \notag \\ 
    &-\Big[m_{12}^2\phi_1^\dagger \phi_2\left(2i\frac{a}{v_s}-2\frac{a^2}{v_s^2}+... \right)+{\rm h.c.}\Big]
    \;,  \label{eq:Vworho}
  \end{align}
  where the mass parameter $m_{12}$ is given by
  \begin{align}
  {m^{ 2}_{12}=-\frac{1}{2}\kappa_{S}v_S^2}\;. \quad
  \end{align}
  For later convenience, we introduce a mass parameter
  \bal
 \tilde{m}_{12}^2=\frac{m_{12}^2}{c_\beta s_\beta} \;,
  \eal
which we use as a model input parameter. Hereafter, we use the following abbreviations when necessary: $s_\theta \equiv \sin\theta$ and $c_\theta \equiv \cos\theta$.
 
  The other mass parameters and the quartic couplings have been replaced with
  \begin{align}
     m_{ii}^{2}&=m_{ii}^{\prime  2}+\frac{1}{2}\kappa_{\phi_i}v_s^2\;, \quad \lambda_i=\lambda'_i \quad(i=1\mbox{-}4)\;.
  \end{align}
  In these relations, we have neglected the contributions proportional to $\kappa_S$ since the size of this coupling constant is $\kappa_S\sim v^2/v_S^2 \ll 1$.  
The quartic couplings $\kappa_{\phi 1}$ and $\kappa_{\phi 2}$ have also been neglected in $\lambda_i$ since the typical size of these couplings is similar to $\kappa_S$. 
    
The physical state of the axion after the electroweak symmetry breaking (EWSB) is obtained by requiring that it is orthogonal to the NG boson $G^0$~(e.g., see the detail in  Refs.~\cite{Srednicki:1985xd, Cox:2023squ}). 
We then have 
\begin{align}
a=\frac{v}{v_a}\Big\{s_{2\beta}(s_\beta z_1-c_\beta z_2)+\frac{v_S}{{v}} \tilde{a}\Big\}\;,
\end{align}
where $v_a$ corresponds to the normalization factor for the axion field $a$, which is defined by
\begin{align}
v_a^2\equiv v_S^2+v^2(\sin2\beta)^2. 
\end{align}
We note that $v_s$ should be larger than the masses of the heavy Higgs bosons. 
Otherwise, the effect of $\rho$ is not negligible, meaning that the effective Lagrangian Eq.~\eqref{eq:Vworho} is not valid.

\subsection{Potential analysis in the broken phase}

We first describe the Higgs potential in the broken phase of the EW symmetry, where ${\expval{S}=v_S/{\sqrt{2}}}$, ${\expval{\phi_1}=v_1/\sqrt{2}}$, and ${\expval{\phi_2}=v_2/\sqrt{2}}$.
By doing so, we define Higgs fields in the mass basis and specify the model input parameters that we will choose.
In the broken phase, all component fields of the Higgs doublets mix with each other.
The axion field $\tilde{a}$ mixes with the CP-odd Higgs bosons $z_i$ due to the portal interaction $\kappa_{S}(S^{2}(\phi_{1}^{\dagger}\phi_{2}^{})+{\rm h.c.})$.
By diagonalizing the mass matrices, one obtains the Higgs bosons in the mass basis.
They are given by linear combinations of the original fields in the interaction basis as follows:
  \begin{equation} \label{eq:massbasis}
    \begin{pmatrix} w_1^\pm\\ w_2^\pm \end{pmatrix}=R_{\beta}
      \begin{pmatrix} G^\pm\\ H^\pm \end{pmatrix}\; , \quad
        \quad
    \begin{pmatrix} h_1\\ h_2 \end{pmatrix} =R_{\alpha}
      \begin{pmatrix} H\\ h \end{pmatrix}\; , \quad
          \begin{pmatrix} z_1\\ z_2\\  \tilde{a} \end{pmatrix}=R^{3\times 3}_\beta R^{3\times 3}_\theta
      \begin{pmatrix} G^0\\ A \\ a \end{pmatrix}\; , \quad
  \end{equation}
  where the orthogonal matrices are defined by  
  \begin{align}
      \label{eq:gammaRots}
    {R}_{x} &=
    \begin{pmatrix}
       \cos x & -\sin{x} \\
       \sin{x} & \cos x \end{pmatrix} \,\,(x=\beta~{\rm or}~\alpha) \,,
       \quad \\
        R^{3\times 3}_\beta&=
  \begin{pmatrix}
    \cos \beta & -\sin\beta &0\\
    \sin\beta & \cos \beta &0\\
    0 & 0 & 1
  \end{pmatrix}\,,  \quad
  R^{3\times 3}_\theta=
  \begin{pmatrix}
    1 & 0 & 0\\
    0&\cos \theta & -\sin \theta \\
    0&\sin \theta & \cos \theta \\
  \end{pmatrix} \,. \label{e:Ogamma1} \,
  \end{align}
  By the rotation with the angle $\beta$ ($\alpha$), the charged Higgs bosons (the CP-even Higgs bosons) are diagonalized. 
  For the CP-odd Higgs boson sector, the mass eigenstates are obtained by the series of rotations with $\beta$ and $\theta$. 
  The mixing angle $\theta$ is given by
  \begin{align}\label{eq:theta}
  \cos\theta=\sqrt{1-\frac{v^2}{v_a^2}s_{2\beta}^2}\;,\quad \sin\theta=\frac{v}{v_a}s_{2\beta} \;.
  \end{align}
  Hence, in the mass basis, the NG bosons ($G^0, G^\pm$) absorbed by the longitudinal mode of the weak gauge bosons, five Higgs bosons ($h, H, A, H^\pm$), and the physical state of the axion ($a$) exist. 
 In the following, we identify $h$ as the discovered Higgs boson with a mass of 125{\rm GeV} and $H, A, H^\pm$ are additional Higgs bosons heavier than $h$. 
 
 The mass eigenvalues for the Higgs bosons and the mixing angle $\alpha$ are calculated as 
  \begin{align}
    m_{H^\pm}^2&=
  \frac{m_{12}^{ 2}}{c_{\beta}s_{\beta}}
  -\frac{1}{2}\lambda_4 v^2,  \\
m_A^2&=\frac{m_{12}^2}{c_\beta s_\beta}\left(1+\frac{ v^2}{v^2_S}s^2_{2\beta}\right)\;, \label{eq:mA}  \\
    m_H^2&=
    ({{\cal M}^2_S})_{11} \cos ^2(\alpha )+({{\cal M}^2_S})_{12} \sin (2 \alpha )+({{\cal M}^2_S})_{22} \sin ^2(\alpha ),
    \\
    m_h^2&=
    ({{\cal M}^2_S})_{11} \sin ^2(\alpha )-2 ({{\cal M}^2_S})_{12} \sin (\alpha ) \cos (\alpha )+({{\cal M}^2_S})_{22} \cos ^2(\alpha),
    \\
    \tan(2\alpha)&=
     \frac{2({{\cal M}^2_S})_{12} }{({{\cal M}^2_S})_{11}-({{\cal M}^2_S})_{22}} ,
  \end{align}
  where the mass matrix $M^2_S$ for the CP even Higgs boson in the basis $(h_1,h_2)$ is given by
  \begin{align}
   ({{\cal M}^2_S})_{11}&=\frac{1}{v_1}(m_{12}^{ 2}v_2+2v_1^3\lambda_1)\;, \\ 
   ({{\cal M}^2_S})_{12}&=-m_{12}^{ 2}+v_1v_2(\lambda_3+\lambda_4)\;,  \\ 
   ({{\cal M}^2_S})_{22}&=\frac{1}{v_2}(m_{12}^{ 2}v_1+2v_2^3\lambda_2)\;.
  \end{align}
  Using these mass formulae, the quartic couplings $\lambda_i$ $(i=1\mbox{-}4)$ are replaced by the masses and mixing angles. 
  Hence, we take the following physical parameter as an input parameter, 
\begin{align}\label{eq:inputs}
  v\;,\quad
  m_h\;,\quad
  m_{H^\pm}\;,\quad m_H\;,\quad 
  \tilde{m}^2_{12}\;,\quad
  s_{\beta-\alpha}\;,\quad
  t_{\beta}\;,\quad {v_a}\;,\quad m_a\;.
\end{align}
The EW VEV and the mass of the Higgs boson are fixed by $v=246{\rm GeV}$ and $m_h=125{\rm GeV}$.

Let us comment on the mass of the axion.
We assume that the axion mass arises from interactions outside the EW sector. Thus,
it is not generated in the EWSB.
As one of the possibilities, one can consider that the axion mass comes from higher-dimensional operators.
Assuming that the higher-dimensional operator which breaks U(1) to $Z_n$ is of the form $(S^n/M_*^{n-4})$, the mass of the axion is estimated as
$m^2_a\sim v_S^{n-2}/M_*^{n-4}$, where $M_*$ is a new physics scale higher than the breaking of the $U(1)$ symmetry.
For instance, in the case of $n=8$, one obtains the keV scale axion as follows: $m_a^2\sim (1{\rm keV})^2(v_s/10^8{\rm GeV})^6(10^{15}{\rm GeV}/M_*)^4$.
Alternatively, if one considers the hidden sector and introduces Yukawa couplings of $S$ with the hidden quarks, the axion mass is generated by the instanton effects of the hidden sector.
In this paper, since we are interested in the phenomenological aspect of axion thermal productions, we treat the axion mass $m_a$ as a free parameter~\footnote{In Type-II the axion mass is generated by the QCD instanton effects as similar to the QCD axion. When we consider the axion of a mass in the {{keV to sub-GeV scale}}, the contributions from QCD instanton are negligible. For Type-I, this contribution does not arise since the axion is anomaly-free.}.

  \subsection{ Axion couplings and decay rates}
  
In this section, we derive various axion couplings with the SM fields and give a formula for the decay rates of the axion. 
\subsubsection{Axion-fermion couplings}
  To derive the axion-fermion couplings, we first give the Yukawa lagrangian. 
  It is written as
  \begin{align}\label{eq:Lyukawam}
    -{\mathcal L}_Y =
    &Y_{u}{\overline Q}_Li\sigma_2\phi^*_uu_R^{}
    +Y_{d}{\overline Q}_L\phi_dd_R^{}
    +Y_{e}{\overline L}_L\phi_e e_R^{}+\text{h.c.},
    \end{align}
  where $\phi_{u,d,e}$ is either $\phi_1$ or $\phi_2$, depending on the charge assignment of the additional $U(1)$.
In this paper, we consider two different charge assignments summarized in Table~\ref{tab:xia}, where ${X_{\phi_f}}$  denotes the $U(1)$ charge for the $\phi_{f}$ $(f=u,d,e)$. 
In Type-I, all fermions couple with $\phi_2$. 
In Type-II, quarks couple with $\phi_2$, but leptons couple with $\phi_1$.
The mass range of the axion that we will pay attention to is the {keV to sub-GeV} scale, where the axion decay into electrons (or photons) is relevant. 
Because of this, although there are other possible charge assignments of the $U(1)$, they give the same phenomenology as in  Type-I or Type-II.

Applying the orthogonal transformation of the scalar fields Eq.~\eqref{eq:massbasis} and the relation \eqref{eq:theta}, one can calculate axion-fermion couplings as 
\begin{align}
{\cal L}_a^Y&=g_{af} ia\bar{f}\gamma_5 f \;, \\  
g_{af}&=2{I_3^f}{X_{\phi_f}}\frac{m_f}{v_a} \;,  \label{eq:gaf}
\end{align}
{where $I_3^f$ is the third component of the weak isospin for a left-handed fermion $f_L$, i.e., $I_3^u=1/2$, $I_3^d=-1/2$ and $I_3^e=-1/2$.}
For example, this yields {$g_{au}=-2c_\beta^2m_u/v_a$, $g_{ad}=-2s_\beta^2m_d/v_a$, and $g_{ae}=-2s_\beta^2m_e/v_a$} for the Type-II. 

  \begin{table}[]
    \caption{The U(1) charge assignment for the Higgs field. 
    \label{tab:xia}}
    \centering
  \begin{tblr}{|c||c|c|c|}
    \hline
                  & ${X_{\phi_u}}$         & ${X_{\phi_d}}$ & ${X_{\phi_e}}$         \\ \hline 
    Type-I     & $X_{2}$ & $X_{2}$& $X_{2}$ \\ 
    Type-II     & $X_{2}$ & $X_{1}$& $X_{1}$ \\ \hline
    \end{tblr}
    \end{table}

  \subsubsection{Axion couplings with gluon and photon}
  We define the axion-gluon coupling and axion-photon coupling by~\cite{ParticleDataGroup:2020ssz, GrillidiCortona:2015jxo}
  \begin{align}
  \mathcal{L}=-\frac{g_{agg}}{4} a G_{\mu\nu}\tilde{G}^{\mu\nu}-\frac{g_{a\gamma\gamma}}{4} a F_{\mu\nu}\tilde{F}^{\mu\nu}\;,
  \end{align}
  where {the} dual field strength tensors are given by $\tilde{G}^{\mu\nu}=(1/2)\epsilon^{\mu\nu\rho \sigma}G_{\rho \sigma}$ and $\tilde{F}^{\mu\nu}=(1/2)\epsilon^{\mu\nu\rho \sigma}F_{\rho \sigma}$. 
  The coupling constants $g_{agg}$ and $g_{a\gamma \gamma}$ can be obtained by calculating 1-loop triangle diagrams for the $agg$ and $a\gamma\gamma$ vertices.  
  In terms of the Passarino-Veltman functions~\cite{Passarino:1978jh}, they are calculated as
\begin{align}
g_{agg}&=\frac{\alpha_s}{\pi}\sum_{q} g_{aqq}m_q C_0(0,0,m_a^2;m_q,m_q,m_q)\;, \\ 
g_{a\gamma\gamma}&=\frac{2\alpha_{\rm em}}{\pi }\sum_f g_{aff} N_c^fQ_f^2 m_f C_0(0,0,m_a^2;m_f,m_f,m_f)\;. 
\end{align}
Using Eq.~\eqref{eq:gaf} and the approximate formula of the $C_0$ function for $m_a \gg m_f$
\begin{align}
C_0(0,0,m_a^2;m_f,m_f,m_f)\simeq -\frac{1}{2m_f^2}\left(1+\frac{1}{12}\frac{m_a^2}{m_f^2}\right)\;,
\end{align}
the above expressions are reduced to 
\begin{align} \label{eq:agg_agamgam}
g_{ag g}=\frac{\alpha_{s}}{\pi v_a}N  \;,\quad
g_{a\gamma \gamma}=\frac{\alpha_{\rm em}}{\pi v_a}(E+E')\;,
\end{align}
where the coefficients $N$ and $E$ ($E'$) correspond to the anomaly (threshold corrections) part, and they are given by 
\begin{align}
N&=-\sum_qI_q {X_{\phi_q}} \;, \\
E&=-\sum_fN_c^f2I_f{X_{\phi_f}}Q_f^2\;,\quad
E'=-\frac{1}{12}\sum_fN_c^f2I_f{X_{\phi_f}}Q_f^2\frac{m_a^2}{m_f^2}\;.
\end{align}
For Type-I and Type-II, they are calculated as 
\begin{align}
N_{\rm Type\mbox{-}I}&= {-}3(X_2-X_2)/2 =0\;, \notag \\
N_{\rm Type\mbox{-}II}&
  = {-}3(X_2-X_1)/2 \;, \\
E_{\rm Type\mbox{-}I}&=3\Big[3\left(\frac{2}{3}\right)^2(2c_\beta^2){-}3\left(-\frac{1}{3}\right)^2(2c_\beta^2){-}\left(-1\right)^2(2c_\beta^2) \Big]=0\;, \\ 
E_{\rm Type\mbox{-}II}&=3\Big[3\left(\frac{2}{3}\right)^2(2c_\beta^2)+3\left(-\frac{1}{3}\right)^2(2s_\beta^2)+\left(-1\right)^2(2s_\beta^2) \Big]=8\;, \label{eq:E_tyII}\\ 
E'_{\rm Type\mbox{-}I}&\simeq-\frac{1}{12}\frac{m_a^2}{m_e^2}2c_{\beta}^2\;, \label{eq:Ep_tyI}\\ 
E'_{\rm Type\mbox{-}II}&\simeq\frac{1}{12}\frac{m_a^2}{m_e^2}2s_{\beta}^2\;, 
\end{align}
where we only included the threshold correction due to electrons and neglected those from other fermions.
In this way, the Type-I gives an anomaly-free axion. 
While the dominant contribution to $g_{a\gamma\gamma}$ comes from the anomaly part in Type-II, the threshold corrections are dominant in Type-I. 
{Although the global U(1) symmetry is not explicitly broken in the Higgs sector in our setup, it is interesting to consider such effects, which could change the axion-photon coupling through mixing between axions and CP-odd Higgs bosons. See Ref.~\cite{Sakurai:2022roq}.}

\subsubsection{Decay rates of the axion}
As mentioned above, we later focus on the axion in the keV to sub-GeV mass scale. 
The relevant decay channels of the axion then are $a\to\gamma \gamma$ and $a\to ee$. 
The decay rates for these processes  are given by
\begin{align}
\Gamma_{a\to \gamma\gamma}&=\frac{m_a^3}{64 \pi}g_{a\gamma\gamma}^2 \;, \\ 
\Gamma_{a\to e^+e^-}&=\frac{m_a{m_e^2}}{8\pi v^2_a}{X_{\phi_e}^2}\sqrt{1-\frac{4m_e^2}{m_a^2}} \;.
\end{align}
\subsection{Potential analysis in symmetric phase}\label{eq:pasymp}
We here describe masses and couplings for scalar particles in the symmetric phase of the EW symmetry since we are interested in the production of the axion from {decay and} scatterings of the heavy Higgs bosons in the thermal equilibrium. 
In other words, we consider the circumstance of ${\expval{S}=v_S/{\sqrt{2}}}$, $\expval{\phi_1}=0$ and $\expval{\phi_2}=0$. 

To make the formalization clear, we use the Higgs basis~\cite{Donoghue:1978cj, Georgi:1978ri, Botella:1994cs, Davidson:2005cw}. 
The Higgs basis is introduced by the transformation
\begin{align} \label{eq:HB}
\begin{pmatrix}
  \phi_1\\
  \phi_2
\end{pmatrix}
=
\begin{pmatrix}
  c_\beta& -s_\beta\\
  s_\beta& c_\beta \\
\end{pmatrix}
\begin{pmatrix}
  H_1\\
  H_2
\end{pmatrix} \;,
\end{align}
where $H_i (i=1,2)$ is the Higgs doubet field in the Higgs basis. 
Applying this transformation in Eq.~\eqref{eq:Vworho}, the Higgs potential in the Higgs basis is given by 
\begin{align}
V &= Y_{1}^{2}(H_1^\dagger H_1) +  Y_{2}^{2}(H_2^\dagger H_2)  
-(Y_{3}^2 H_1^\dagger H_2+{\rm h.c.})
 \nonumber \\
 &+\frac{Z_1}{2}(H_1^\dagger H_1)^2 + \frac{Z_2}{2}(H_2^\dagger H_2)^2 
+ Z_3 (H_1^\dagger H_1)( H_2^\dagger H_2) 
+ Z_4 (H_1^\dagger H_2)( H_2^\dagger H_1) \;, \notag \\
&+\Big\{Z_5 (H_1^\dagger H_2)^2+Z_6 (H_1^\dagger H_1)^2(H_1^\dagger H_2)+Z_7 (H_2^\dagger H_2)^2(H_1^\dagger H_2) +{\rm h.c.} \Big\}   \notag \\
&-\Big[m_{12}^2 \left(2i\frac{a}{v_s}-2\frac{a^2}{v_s^2}+... \right)
(c_\beta^2H_1^\dagger H_2-s_\beta^2 H_2^\dagger H_1+\frac{s_{2\beta}}{2}\{ H_1^\dagger H_1-H_2^\dagger H_2\} )
+{\rm h.c.}\Big] \;. \label{eq:VHB}
\end{align}
The relation between $(m^2_{ii},\lambda_j)$ and $(Y_i,Z_j)$ are presented in Appendix~\ref{ap:HBparameter}.
At this stage, the Higgs fields have not been diagonalized. The mass matrix is given by
\begin{align}
{\cal M}^{{2}}_{ H_{1}H_{2}}=
\begin{pmatrix}
  Y_1^2&-Y_3^2 \\
  -Y_3^2&Y_2^2 
\end{pmatrix}
\end{align}
We can write this potential in terms of the mass basis where the mass terms of the two Higgs doublet fields are diagonalized.
The transformation matrix is given by
\begin{align} \label{eq:HBMB}
\begin{pmatrix}
  H_1\\
  H_2
\end{pmatrix}
=
R_\omega
\begin{pmatrix}
  H^\prime_1\\
  H^\prime_2
\end{pmatrix}\;,
\end{align}
where we denote the Higgs fields in the mass basis by $(H_1^\prime,H_2^\prime)$ and  
the mixing angle $\omega$ is given by 
\begin{align}
\tan(2\omega)=\frac{2Y_3^2}{Y_2^2-Y_1^2}
=-\frac{s_{\beta-\alpha}c_{\beta-\alpha}(m_H^2-m_h^2)}{s_{\beta-\alpha}c_{\beta-\alpha}(m_H^2-m_h^2)\cot2\beta+\tilde{m}^2_{12}}.
\end{align}
Thus, the original basis $(\phi_1,\phi_2)$ is related to the mass basis by $(\phi_1,\phi_2)^T=R_{\beta} R_{\omega} (H'_1,H'_2)^T$. 
The potential is easily obtained by the following replacement,
\begin{align}
&H_{1,2}\to  H^\prime_{1,2} \;,
\\
&Y_{1,2}\to Y_{1,2}^\prime\;,\quad Y_{3}\to 0\;,\quad Z_{i}\to Z_{i}^\prime \  \mbox{$(i=1,...,7)$}
\end{align}
in Eq.~\eqref{eq:VHB}. 
Concrete expressions of $(Y^\prime_i,Z_j^\prime)$ are presented in Appendix~\ref{ap:mass coupling in MB SP}, using the model input parameters Eq.~\eqref{eq:inputs}. 
We parameterize the component fields of $H'_{1,2}$ as
\begin{eqnarray} 
  H_k' = \frac{1}{\sqrt{2}}
  \begin{pmatrix} \sqrt{2}\, w_k^{\prime+} \\  h'_k
    + i z'_k  \end{pmatrix} \; (k = 1,2)\;.\quad
  \label{eq:fieldHBMass}
\end{eqnarray}
Since in this basis, the mass matrix for the two Higgs doublet fields are diagonalized, the masses for $H^\prime_1$ and $H^\prime_2$
are given by $m^2_{H'_1}=Y_1^{\prime 2}\;,m^2_{H'_2}=Y_2^{\prime 2}$, where the subscripts denote $H'_1=h'_1,z'_1$ or $\omega^{'\pm}_2$ and $H'_2=h'_2,z'_2$ or $\omega^{'\pm}_2$.  
By expanding in $c_{\beta-\alpha}$ assuming $|c_{\beta-\alpha}| \ll 1$, the masses and the mixing angle are approximately given by
\begin{align} \label{eq:mass_in_MBSP}
m^2_{H^\prime_1}&\simeq -\frac{m_h^2}{2}-c_{\beta-\alpha}^2\frac{1}{4 \tilde{m}^2_{12}}(m_H^2-m_h^2)(2\tilde{m}^2_{12}+m_H^2-m_h^2)\;,\\ \label{eq:MH2}
m^2_{H^\prime_2}&\simeq -\frac{m_h^2}{2}+ \tilde{m}^2_{12}-c_{\beta-\alpha}\frac{t_\beta^2-1}{2t_\beta}(m_H^2-m_h^2)\;,\\
\tan2\omega&\simeq -\frac{1}{\tilde{m}^2_{12}}(m_H^2-m_h^2)c_{\beta-\alpha}\;.
\end{align}
We note that the axion field $\tilde{a}$ does not mix with the CP-odd component fields of $H_{1,2}$ in the symmetric phase. 

In the mass basis of the symmetric phase, we have the following trilinear and quartic couplings involving the axion,
\begin{align} \label{eq:coupH1H2a}
\lambda_{ah^\prime_1z_2^\prime }&={-}\frac{\tilde{m}^2_{12}}{v_S}s_{2\beta} \;,\quad 
{-}\lambda_{az^\prime_1h_2^\prime }=i \lambda_{aw^{\prime +}_1w^{\prime -}_2 }
={-}i \lambda_{aw^{\prime +}_2w^{\prime -}_1 }=\lambda_{ah^\prime_1z_2^\prime }\;, \\
\lambda_{H_1^\prime H_1^\prime aa}&= -4 s_{2\beta+2\omega}c_\beta s_\beta \frac{\tilde{m}^2_{12}}{v^2_S}\;,\quad 
-\lambda_{H_2^\prime H_2^\prime aa}=\cot(2\beta+2\omega) \lambda_{H_1^\prime H_2^\prime aa}= \lambda_{H_1^\prime H_1^\prime aa}\;, 
\end{align}
where for the quartic couplings, the superscripts denote $H'_{i}H'_{i}=h'_i h'_i,\;z'_i z'_i,\;\omega^{\prime +}_i \omega^{\prime -}_i,\;$ $(i=1,2)$ and $H'_{1} H'_{2}=h'_1h'_2,\;z'_1z'_2,\; {\omega^{\prime +}_1\omega^{\prime -}_2,\; \omega^{\prime +}_2\omega^{\prime -}_1}$. 
Through these scalar couplings, thermal production of axions from the heavy Higgs bosons can occur, which will be discussed in the next section.

\section{Axion productions}\label{sec:axion productions}

In this section, we evaluate the relic abundance of the axion from the heavy Higgs boson decays (Sec.~\ref{sec:production from decay}) and scatterings (Sec.~\ref{sec:production from scattering}). 
For this analysis, we assume the following conditions are met:\footnote{
In the parameter region of our interest,  axions do not reach thermal equilibrium.
}
\begin{itemize}
  \item  Heavy Higgs bosons reach thermal equilibrium in the thermal history of the Universe. 
  Under this assumption, heavy Higgs bosons are abundantly produced from the thermal bath when the Universe's temperature exceeds their mass.
  {Therefore, we assume a reheating temperature higher than the mass of heavy Higgs bosons,  $T_R>m_{H'_2}$, to ensure their production from the thermal bath.}
  \item The mass of the heavy Higgs bosons significantly exceeds the EW VEV, implying their efficient production before EWSB.
  As can be seen from Eq.~\eqref{eq:MH2}, the heavy Higgs bosons acquire a mass from the U(1) breaking term~$m_{12}^2\phi_1^\dagger \phi_2$. 
  This assumption is consistent with the collider constraints on the heavy Higgs bosons, as discussed in Sec.~\ref{sec:numerical results}. 
\end{itemize}
In the symmetric phase, particles in the thermal bath receive thermal corrections to their masses. 
In the following calculations of the axion abundance, we incorporate the thermal mass corrections. 
The thermal mass effect is particularly important for $H'_1$ since it has a tachyonic mass unless thermal corrections are included, as can be seen in Eq.~\eqref{eq:mass_in_MBSP}. 
We denote the thermal corrected masses by the capital letter, e.g., $M_{H'_1}, M_{H'_2}$ for the $H'_1$ and $H'_2$, respectively. These masses depend on the temperature $T$, and the analytical formulae are given in Appendix~\ref{ap:thermal mass}.

Here, we comment on the axion production from the SM particles. 
As mentioned above, we assume that the reheating temperature is higher than the heavy Higgs boson mass. 
In this symmetric phase,
axion production from the SM particles is negligible for several reasons.
One is that both anomalous axion-gauge boson couplings and the axion derivative couplings with fermions are absent in the symmetric phase.
In addition, the axion does not appear in the Yukawa Lagrangian since it has no mixing with the CP odd Higgs.
Thus, there is no axion production through gauge couplings $g_i$ $(i=1,2,3)$ or  Yukawa couplings in the symmetric phase. 
The axion interactions only appear in the Higgs potential as given in Eq.~\eqref{eq:VHB}. 
The second reason is that the axion trilinear couplings necessarily involve the additional Higgs bosons (see Eq.~\eqref{eq:coupH1H2a}). 
Thus, the final states of the SM-like Higgs boson decay should contain the additional Higgs bosons in addition to axion, which is kinematically forbidden. 
On the other hand, scattering with the SM-like Higgs boson can happen. 
We will discuss that they do not become main axion production channels in Sec.~\ref{sec:production from scattering}. 
Therefore, in our setup, the axion is mainly produced from heavy Higgs bosons. 

In the broken phase, the axion couples to SM particles by mixing with CP-odd scalar states.
This enables the SM-like Higgs boson to decay into an axion and a weak gauge boson.
In addition, the axion-top quark coupling through this mixing allows axion production via top-quark scattering.
If $T_R$ is higher than the temperature of EWSB, these processes would be especially important ones among axion production processes in the broken phase since thermal axion production through the renormalizable interactions is IR dominant (concrete examples can be seen in Sec.~\ref{sec:production from decay} and Sec.~\ref{sec:production from scattering}). 
{The SM-like Higgs boson decay and the top quark scatterings are an irreducible contribution to the axion production under the assumption that all the SM particles are thermalized. 
When heavy Higgs bosons are not thermalized, these contributions dominate axion production.
 }
In Sec.~\ref{sec:Axion_productions_in_the_broken_phase}, we discuss axion productions from the SM-like Higgs boson decay and the Top Yukawa coupling. We then compare the amount of the produced axion with the one from the heavy Higgs boson decay.

\subsection{Productions from the heavy Higgs boson decays } \label{sec:production from decay}

Under the above-mentioned assumptions, decays of the heavy Higgs bosons producing axions occur in the symmetric phase.
The relevant process for axion production from heavy Higgs boson decays is
\bal \label{eq:process_decay}
H'_2\to H'_1 a\;,
\eal
which includes $h'_2\to z_1' a$, $z'_2\to h_1' a$, and $w^{\prime \pm}_2\to w^{\prime \pm}_1 a$. 
All these processes occur through the trilinear couplings presented in Eq.~\eqref{eq:coupH1H2a}, which arise from the portal interaction ${\kappa_S}S^2 \phi_1^\dagger\phi_2$.
It should be noted that these are the only couplings that generate decay processes into axions at the tree level.
Decay processes involving gauge bosons in the final state, e.g., $h'_2\to a Z$, do not occur at the tree level because gauge-scalar-scalar type couplings only arise after the Higgs boson fields develop their VEVs.

Before formulating the yield of axions from heavy Higgs boson decays, we estimate the temperature dependence of this process. 
From naive dimensional analysis, the time derivative of the yield can be expressed as
\bal
S\dot{Y}_{a}^{D}\simeq \int d\Pi_{H'_2}m_{H'_2}f_{H'_2}\Gamma_{H'_2\to H'_1 a}
\sim T^2 m_{H'_2}f_{H'_2}\Gamma_{H'_2\to H'_1 a} \;. 
\eal 
where $\Gamma_{H'_2\to H'_1 a}$ is the decay rate of heavy Higgs bosons, which will be presented below shortly. 
Here {$S$ denotes the entropy density. }
The yield ${Y}_{a}^{D}$ is defined as $Y_a^D=n^D_a/S$. 
The superscript $D$ indicates that only the decay processes of the heavy Higgs bosons are considered in the evaluation of the Boltzmann equation. 
$f_{H'_2}$ denotes the distribution function of ${H'_2}$ and the Maxwell Boltzmann distribution $f_{H'_2}=\exp(-E_{H'_2}/T)$ is used. 
Using $ \dot{T}\simeq - H T$, $H\sim T^2/m_{\rm pl}$, and $S\sim T^3$, the yield of axions produced over the Hubble time at temperature $T$ can be estimated as
\bal \label{eq:yielddecap}
Y_a^D\sim \frac{m_{\rm pl} m_{H'_2}}{T^3}e^{-m_{H'_2}/T}\Gamma_{H'_2\to H'_1 a} \;,
\eal
where $m_{\rm pl}$ denotes the Planck mass. 
Thus, the yield is maximized at $T\sim m_{H'_2}$ and  
the axion production is suppressed at $T\ll m_{H'_2}$ due to the Boltzmann suppression factor.  

We then describe the yield parameter obtained by rigorously solving the Boltzmann equation. 
The key ingredient is the decay rate of the heavy Higgs bosons, which is given by
\bal \label{eq:h2ptoaz1}
\Gamma (h^\prime_2 \to a z^\prime_1) 
\simeq
\left(\frac{\tilde{m}^2_{12}}{v_S} s_{2\beta}\right)^2\frac{1}{16\pi M_{H'_2}}\Bigg(1-\frac{M^2_{H'_1}}{M^2_{H'_2}} \Bigg) \;,
\eal
{where $M_{H'_1}$ and $M_{H'_2}$ are the thermal corrected masses.}
As given in Eq.~\eqref{eq:coupH1H2a}, the relevant couplings for axion production are common, and the masses of scalar particles are degenerate  in the symmetric phase, so that other production processes are written as
\bal
\Gamma(z^\prime_2 \to a h^\prime_1 )=\Gamma(w^{\prime \pm}_2 \to a w^{\prime \pm}_1 )
=\Gamma (h^\prime_2 \to a z^\prime_1) \;. 
\eal
This yields the sum of the decay rates for the heavy Higgs bosons as $\Gamma_{H'_2\to a H'_1}=4\Gamma(h^\prime_2 \to a z^\prime_1)$. 
  {In the following analysis, we consider that the axion is not thermalized.}
Thus, the yield for the decays of heavy Higgs bosons can be expressed as~\cite{Hall:2009bx}, 
\bal \label{eq:Yh2toh1a}
 {Y^D_a\simeq4\frac{45 m_{\rm pl}}{1.66\cdot 4\pi^4}\frac{M_{H'_2}^2\Gamma_{h^\prime_2 \to a z^\prime_1}}{{\tilde{m}_{12}^4} } \int^{x_{\rm max.}}_{x_{\rm min.}}\frac{x^3}{g^{3/2}_\ast(x)}K_1\left(\frac{M_{H'_2} }{{\tilde{m}_{12}}}x\right)d x} \;,
\eal
where the prefactor four accounts for all decay processes, i.e., $h^\prime_2 \to a z^\prime_1,\ z^\prime_2 \to a h^\prime_1, w^{\prime \pm}_2 \to a w^{\prime \pm}_1$, and the fraction parameter $x\equiv {\tilde{m}_{12}}/T$ is introduced~\footnote{{We have used that the initial degree of freedom of $H_{2}'$ is $g_{H'_2}=1$ in Eq.~\eqref{eq:Yh2toh1a}. The same is true for Eq.~\eqref{eq:Rx}.}}. 
{ We simply set the lower limit of the integration by $x_{\rm min.}=0$ since the process is IR dominant. 
The upper limit is taken to be $x_{\rm max.}=\tilde{m}_{12}/T_0$.}
The lowest temperature $T_0$ corresponds to the temperature at which the mass of $H'_1$ becomes zero. 
The effective degree of freedom $g_\ast$ is calculated using the data given in Ref.~\cite{Saikawa:2020swg}.
The function $K_1$ is the first modified Bessel Function of the second kind. 
In deriving Eq.~\eqref{eq:Yh2toh1a}, we have assumed that the initial abundance of $a$ is negligible, 
and so, the back reaction of Eq.~\eqref{eq:process_decay} is not included in our computation. 
We have also neglected the effect of Bose condensation and set $1-f_{H_1^\prime}\simeq 1$. 
{We emphasize that Eq.~\eqref{eq:Yh2toh1a} is a general formula for thermal {freeze-in} production from the heavy Higgs boson decays. Thus, one can apply this formula to the original DFSZ QCD axion model.} 

\begin{figure}[t]
  \centering
  \includegraphics[scale=0.7]{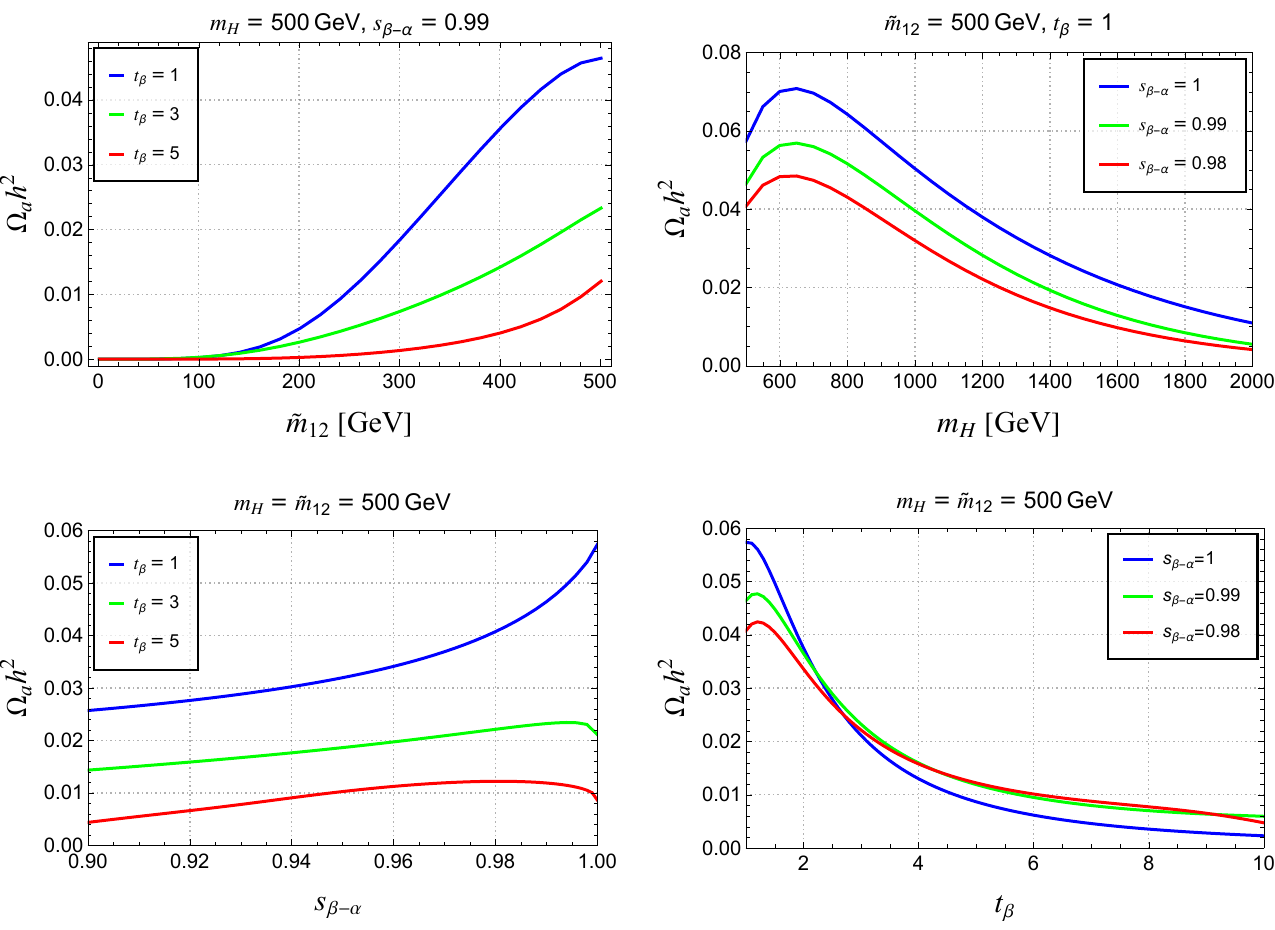}
  \caption{The relic abundance of the axion $\Omega_a h^2$ evaluated from Heavy Higgs boson decays. }
  \label{fig:omegaD}
\end{figure}

Let us describe the model parameter dependence of the yield of axions produced from heavy boson decays. 
Substituting the decay rate Eq.~\eqref{eq:h2ptoaz1} into the approximate formula of the yield Eq.~\eqref{eq:yielddecap}, one can see that the yield $Y_{a}^{\rm D}$ scales as
\bal \label{eq:YaD scaling}
{Y_{a}^{\rm D}\propto \frac{\tilde{m}^4_{12}}{M^{\prime 3}_{H_2}} s_{2\beta}^2}, 
\eal
where we assume $T\sim M'_{H_2} $. 
As can be seen from this scaling law, the relic density of $a$ increases (decreases) as ${\tilde{m}_{12}}$ ($t_\beta$) increases. 
The dependence on $s_{\beta-\alpha}$ and $m_H$ appears through $M_{H_2}'$.
The mass $M_{H_2}'$ becomes large with larger $m_H$ (or smaller $s_{\beta-\alpha}$), so that the relic density shrinks in such a case. 
We also show numerical results of {the density parameter of the axion} $\Omega_a h^2$ for Type-II in Fig.~\ref{fig:omegaD}. 
{This quantity is evaluated by $\Omega_a=\rho_a/\rho_{c0}\simeq (S_0/\rho_{c0})m_aY^D_a\simeq 274.3(m_a/1{\rm keV})h^{-2}Y_a^D$, where in the last equality we have substituted the entropy and the critical density at present $S_0=2.89\times 10^{3}{\rm cm}^{-3}$~\cite{ParticleDataGroup:2024cfk} and $\rho_{c0}=1.05\times 10^{-5} h^2 {\rm GeV} {\rm cm}^{-3}$~\cite{ParticleDataGroup:2024cfk}.}
{In the figure, } we chose the benchmark point,  $m_a=1{\rm keV},v_a=10^{10}{\rm GeV},m_H=m_{H^\pm}={\tilde{m}_{12}}=500{\rm GeV}, t_\beta=1,$ and $s_{\beta-\alpha}=0.99$. 
For this benchmark point, one of the four input parameters - ${\tilde{m}_{12}}$, $m_H$, $t_\beta$, or $s_{\beta-\alpha}$ - is varied in each respective panel.
These results follow the behavior that can be understood from the above-scaling law of the yield $Y_{a}^D$. 
{Because of the difference between Type-I and Type-II in the Yukawa sector, thermal corrections to the masses of the Higgs bosons in Type-I are different from those in Type-II.}
We note, however, that almost the same results are obtained if we calculate $\Omega_a h^2$ in Type-I since the production from the Higgs sector is common.  

\subsection{Production from heavy Higgs boson scattering}\label{sec:production from scattering}
We now turn to axion production from $2\to 2$ scattering processes involving heavy Higgs bosons in the initial state.
There are two possibilities for this type of axion production: single production and double production.
The latter is consistently suppressed by a factor of $1/v_a^2$ at the amplitude level, as the amplitude necessarily involves either the quartic coupling $\lambda_{H'_iH'_j aa}$ or the square of the trilinear coupling $\lambda_{H'_iH'_j a}$.
Double axion production is negligible in the parameter region of interest (we consistently assume $v_a\gtrsim10^8{\rm GeV}$ throughout this paper)
\footnote{To verify the evaluations of the yield parameter presented in Sec.\ref{sec:production from scattering}, we have calculated the yield of $h'_1 h'_1\to aa$ and cross-checked it with Ref.\cite{Heeba:2018wtf}.}.
Therefore, single-axon production is the more significant process.

\begin{figure}[t]
  \begin{center}
  \includegraphics[scale=0.28]{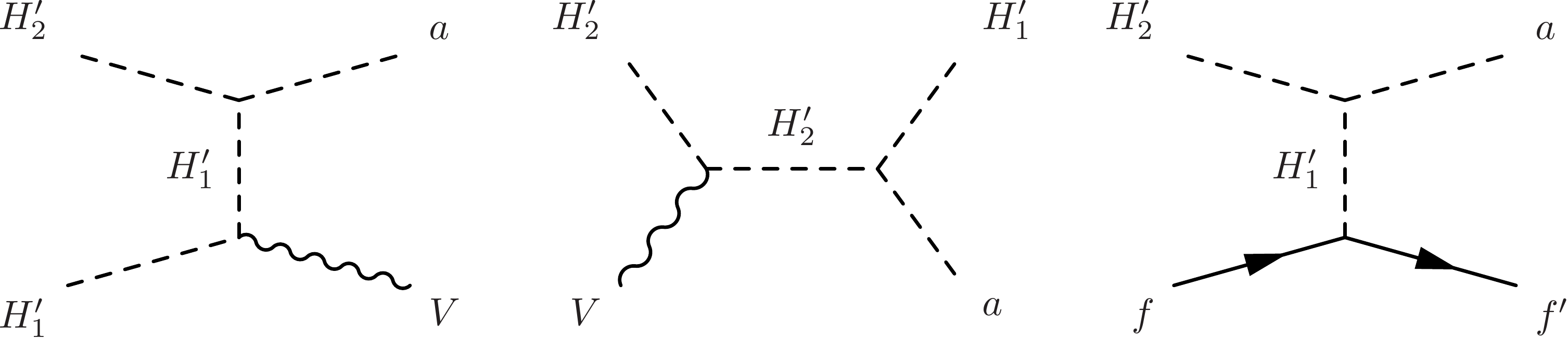}\\
  \end {center}
  \quad\quad~
  (1): $H'_2 H'_1\to a V$ 
    ~~\quad\quad\quad  \quad\quad
  (2): $H'_2 V \to a H'_1$
  ~\quad\quad\quad\quad\quad
  (3): $H'_2 f\to f a$
  \caption{The Feynman diagrams for the single axion production from the heavy Higgs boson scatterings. }
  \label{fig:diagram_scattering}
\end{figure}

At the tree level, we have the following processes,
\bal \label{eq:scaterings}
 &H_{2}^\prime H_1^\prime\to V a\;, \notag  \\
 &H_2^\prime V  \to H_{1}^\prime a\;, \notag  \\
 &H^\prime_{2} f\to f a \;, 
\eal
where $H'_{i}= h_i',\;z_i',\;\omega^{\prime \pm}_i$ $(i=1,2)$ and $V$ represents weak gauge bosons.
The corresponding Feynman diagrams are shown in Fig.~\ref{fig:diagram_scattering}.  
We note that the process $H_{2}^\prime H_1^\prime\to V a$ $(H_2^\prime V  \to H_{1}^\prime a)$ has the corresponding u-channel (t-channel) diagram, {which is not included in the figure but included in our computations}.

We first present the temperature dependence of axion production from the 2-to-2 scattering process $H_2' X_1\to X_2 a$.
Neglecting the effects of Bose enhancement and Pauli blocking and assuming that the initial axion abundance satisfies $f_a\simeq 0$, the yield parameter from the scattering $Y^S$ is given by:
\begin{equation}
S\dot{Y}_a^{H_2' X_1\to X_2 a}\simeq g_{X_1}\int d\Pi_{H_2'} d\Pi_{X_1} \kappa^{1/2}(s,M_{2}^{\prime 2},M_{X_1}^2)\sigma_{H_2' X_1\to X_2 a} f_{H_2'} f_{X_1} ,
\end{equation}
where $(m_{X_{1}},g_{X_{1}})$ are the mass and internal degrees of freedom of $X_{1}$, respectively. The Källén function is defined as:
\begin{equation}
\kappa(a,b,c)=a^2+b^2+c^2-2ab-2bc-2ca.
\end{equation}
Similar to the estimation of the temperature dependence of axion production from decay, naive dimensional analysis yields:
\bal
Y^{H_2' X_1\to X_2 a}_a\sim  g_{X_1}\frac{m_{\rm pl}}{T^3}e^{-m_{H'_2/T}}\frac{{\tilde{m}_{12}^4}}{{v_a^2}}
\eal
where we assume the cross section scales as $\sigma_{H_2' X_1\to X_2 a}\sim \frac{{\tilde{m}_{12}^4}}{{v_a^2}}\frac{1}{T^4} $.
We again observe that the yield is maximized at $T\sim m'_{H_2}$, indicating that the 2-to-2 scattering process is also an IR-dominated process.

We then give the analytical formula for the contributions from the $2\to 2$ scattering process $H_2' X_1\to X_2 a$ to relic density of $a$. 
Following Ref.~\cite{Langhoff:2022bij},
we obtain the formula of the yield for the 2-to-2 scattering process $H_2' X_1\to X_2 a$
as 
\bal
Y^{H_2' X_1\to X_2 a}_a&\simeq
\frac{45 m_{\rm pl}}{1.66\cdot 4\pi^4 }
\int^{ x_{\rm max.}}_{x_{\rm min.}}\frac{x^3}{g^{3/2}_\ast(x){\tilde{m}_{12}^4}}R(x)d x \;, \\ \label{eq:Rx}
R(x)&=\frac{g_{X_1}}{32\pi^4}\int^{\infty}_{(M_{X_1}+M_{X_2})^2}ds
\frac{1}{\sqrt{s}}
K_1\left(\frac{\sqrt{s}}{{\tilde{m}_{12}}}x\right)\kappa_{H'_{1}X_{1}}\sigma_{H'_2X_1\to X_2a} 
\;,
\eal
with $s$ being Mandelstam variable and $M_{X_1,X_2}$ are the thermal corrected masses for $X_1,X_2$.
We here assume that the initial state particles $X_1$ and $X_2$  follow Maxwell-Boltzmann distribution, i.e., $f_{H'_{2}} f_{X_1}\simeq \exp(-(E_{H'_{2}}+E_1)/T)$ with $E_{1}$ being the energy of $X_{1}$. 
The interaction rate $R(x)$ involves the kinematical varialble $\kappa_{xy}$, which is defined by $\kappa_{xy}=\kappa(s,M_x^2,M_y^2)$. 
{The integration limit is common with $Y_a^D$, i.e., Eq.~\eqref{eq:Yh2toh1a}.}
In the following, we give the analytical formulae of the cross sections for each process in order.

The process $H_1^\prime H_2^\prime \to V a$ involves the following fourteen processces in total, $z_2^\prime z^{\prime}_1 \to Z a $, $h_2'h_1' \to a Z$, $h_2^{\pm\prime} h^\prime_1 \to a W$, $h_2^{\pm\prime}z_1'  \to a W$, $h_1^{\pm\prime} h_2' \to a W$, $h_1^{\pm\prime} z_2' \to a W$, $h_2^{\pm\prime} h_1^{\pm\prime} \to a Z$, $h_2^{\pm\prime}$, and $h_1^{\pm\prime} \to a \gamma$. 
We omit the thermal corrections to the masses for the final state particle. 
By making use of the fact that all gauge bosons are massless in this calculation, one can see that all amplitude is proportional to that of $z_2^\prime z^{\prime}_1 \to Z a $:
\bal \label{eq:xsH1H2Za}
{\cal M}_{h_2'h_1' \to a Z} &= {\cal M}_{z_2' z_1' \to a Z}, \\
-{\cal M}_{h_2^{\pm\prime} h^\prime_1 \to a W} &={\cal M}_{h^{\pm\prime}_1 h'_2 \to a W}= 2c_W{\cal M}_{z_2' z'_1 \to a Z}, \\
{\cal M}_{h^{\pm\prime}_2 z_1' \to a W} &=-{\cal M}_{h^{\pm\prime}_1 z^\prime_2 \to a W}= 2ic_W{\cal M}_{z'_2 z'_1 \to a Z}, \\
{\cal M}_{h^{\pm\prime}_2 h^{\mp\prime}_1 \to a Z} &=-2c_{2W}{\cal M}_{z'_2 z'_1 \to a Z}, \\
{\cal M}_{h^{\pm\prime}_2 h^{\mp\prime}_1 \to a \gamma} &=-2s_{2W}{\cal M}_{z'_2 z'_1 \to a Z}.
\eal
Thus, the cross section for ${H_1^\prime H_2^\prime \to V a }$ is given by $\sigma_{h_2'h_1' \to a Z}$ as 
\bal \label{eq:xsH1H2Zaap}
\sigma_{H_1^\prime H_2^\prime \to V a}={(10+32 c_W^2)}\sigma_{z_2' z_1' \to Z a}
\eal
with~\footnote{ We checked that consistency of the squared amplitude $|{\cal M}_{z_2^\prime z_1^{\prime}\to Z a}|^2$ with the computation by {\tt FormCalc}~\cite{Hahn:1998yk}.}
\bal 
\sigma_{z_2^\prime z_1^{\prime}\to Z a}
 &=
\frac{4\lambda_{ah_1'z_2'}^2}{16\pi\kappa_{H'_1H'_2}s}\frac{m_Z^2}{v^2}
\Big[-2\sqrt{\kappa_{H'_1H'_2 }} \notag \\
&+(M^2_{H'_1}+M^2_{H'_2}-s)\log{\left(\frac{M^2_{H'_1}+M^2_{H'_2}-s+\sqrt{\kappa_{H'_1H'_2}}}{M^2_{H'_1}+M^2_{H'_2}-s-\sqrt{\kappa_{H'_1H'_2}}}\right)}\Big]\;.
\eal

Another process that contributes to the axion productions from the scattering is $H'_2 V \to H'_1 a$. 
There are fourteen processes in total. 
All the amplitude can be evaluated by utilizing the crossing symmetry from the amplitudes of $H_2'V\to H_1'a$. 
Hence, similar to $H'_2 V \to H'_1 a$, summing  possible Higgs states yields 
 \bal
 \sigma_{H'_2 V \to H'_1 a }= (10+32c_W^2)\sigma_{z'_2 Z\to z'_1 a}\;. 
 \eal
The cross section for $z'_2 Z\to z'_1 a$  is calculated as
\bal\label{eq:sigmaAZtoG0a} 
\sigma_{z'_2 Z\to z'_1 a}=\frac{1}{8\pi\kappa_{H_2'Z}}\frac{\lambda^2_{ah'_1z'_2}m_Z^2}{v^2(s-M_{H_2'}^2)}
\left[2M_{H'_1}^2-2s-(M_{H'_1}^2+s)\log{\left(\frac{M_{H'_1}^2}{s}\right)}\right]\;.
\eal
Since we identify the $Z$ boson as a massless particle in the symmetric phase, The kinematical factor $\kappa_{z'_2 Z}$ is given by $\kappa_{z'_2 Z}=(s-M_{H_2'}^2)^2$. 
As can be seen in Eq.~\eqref{eq:sigmaAZtoG0a}, the cross section does not converge if the Mandelstam variable $s$ takes the minimal value $s_{\rm min}=M_{H_2'}^2$
\footnote{In the reaction rate $R(x)_{z'_2 Z\to z'_1 a}$, the kinematical factor $\kappa_{H'_2 Z}$ in the denominator of Eq.~\eqref{eq:sigmaAZtoG0a} cancels with the additional one.}. 
Hence, in the following numerical evaluation, we evaluate $\sigma_{z'_2 Z\to z'_1 a}$ by applying the Bright-Wigner propagator in the amplitudes, i.e., $(s-M_{H'_i}^2)^{-1}\to(s-M_{H'_i}^2+iM_{H'_i}\Gamma_{H'_i}^{\rm tot} )^{-1}$ $(i=1,2)$.

The remaining 2-to-2 scattering process for the axion production is $H'_2 f\to  f' a$. 
Considering only the top Yukawa coupling, the cross section for $H'_2 f\to  f' a$ is reduced as
\bal\label{eq:sigmaH2ftofa}
\sigma_{H'_2 f\to  f' a}&\simeq \sigma_{h_2' t \to t a}+\sigma_{z_2' t \to t a}+\sigma_{h^{-\prime}_2 t \to b a}+\sigma_{h^{+\prime}_2 b \to t a} \notag \\
&=4\sigma_{z_2' t \to t a}\;,
\eal
with $\sigma_{z_2' t \to t a}$ being 
\bal
\sigma_{z_2' t \to t a}&=
\frac{\lambda^2_{ah'_1z'_2}}{16\pi \kappa_{H_2't}}\frac{m_t^2}{v^2(s+M^2_{H_1'}-M^2_{H_2'})}
\Bigg[M^2_{H_2'}-s \notag \\
&-(M^2_{H_1'}-M^2_{H_2'}+s)\log{\left(\frac{M^2_{H_1'}}{M^2_{H_1'}-M^2_{H_2'}+s}\right)}\Bigg]\;,
\eal
In the second equality of Eq.\eqref{eq:sigmaH2ftofa}, we have assumed that the external fermions are massless.

\begin{figure}[t]
  \centering
  \includegraphics[scale=0.8]{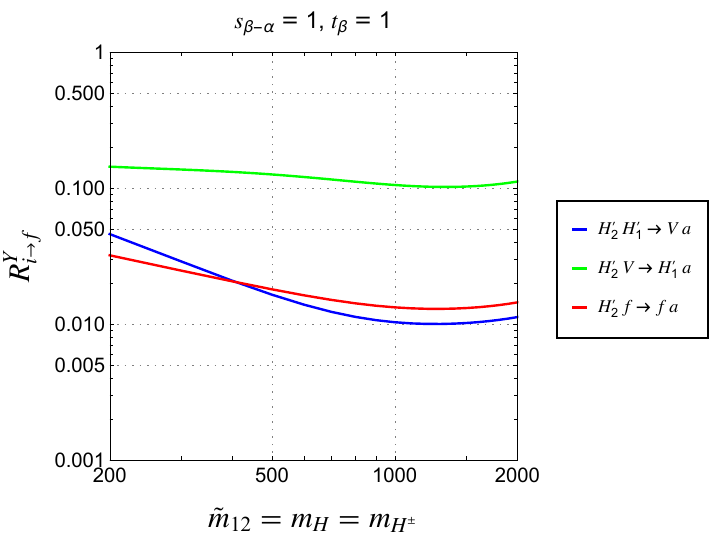}
   \caption{
    The ratio $R^Y_{H_2'X_1\to X_2 a} \equiv Y^{H_2'X_1\to X_2 a}_a/Y^{H_2^\prime \to H_1^\prime a}_a$ is shown by a function of the mass parameter $\tilde{m}_{12}$. 
   }
  \label{FIG:ratio of scattering and decay}
 \end{figure}

 We compare the relic density produced by the 2-to-2 scatterings with that produced by the decay $ H_2^\prime \to H_1^\prime a$. 
{To do this, we define the ratio between the yield parameter from the decay and that from the scattering, i.e., $R^Y_{H_2'X_1\to X_2 a} \equiv Y^{H_2'X_1\to X_2 a}_a/Y^{H_2^\prime \to H_1^\prime a}_a$. }
 {The ratio $R^Y_{H_2'X_1\to X_2 a} $} is shown  as a function of the mass parameter $\tilde{m}_{12}$ in Fig.~\ref{FIG:ratio of scattering and decay}, where the model parameter is fixed by ${v_a}=10^{10}{\rm GeV}$, $s_{\beta-\alpha}=t_\beta=1$, and $m_H=m_{H^\pm}$. 
 {It} is not highly dependent on $\tilde{m}_{12}$.
 {We note that the $v_a$ dependence is canceled out in the ratio $R^Y_{H_2'X_1\to X_2 a}$.}
 Since the $\lambda_{a h_1'z_2'}^2$ appearing in each $Y^{H_2'X_1\to X_2 a}_a$ and $Y^{H_2^\prime \to H_1^\prime a}_a$ cancels in $R^Y_{H_2'X_1\to X_2 a}$, this slight depndence on ${\tilde{m}_{12}}$ comes from the thermal corrected masses $M_{H'_{1,2}}$. 
 In the regime of ${\tilde{m}_{12}}\sim m_h$, $Y^{H_2^\prime \to H_1^\prime a}_a$ is small due to the phase space suppression, so that the ratio slightly increases.
 One can also see that the total amount of the axion produced from $Y^{H_2'V\to H_1' a}_a$, which is a dominant contribution in the scattering process, is small by a factor of ${\mathcal O}(10^{-1})$, compared with the production from the decays.
Therefore, the axion thermal productions from the heavy Higgs bosons are dominated by the decay process $H'_2\to H_1'a$. 
 
{Before we close this section, let us mention the productions from scatterings of the SM particles in the symmetric phase. 
Though the axion trilinear couplings Eq.~\eqref{eq:coupH1H2a}, scattering processes, e.g., $H'_1 V \to H'_2 a$, $H'_1 f \to \bar{f} a$, are also realized in addition to the heavy Higgs boson scatterings. 
We omit them in our analysis~\footnote{{These processes are evaluated in Ref.~\cite{DEramo:2021lgb}.}}. 
The reason is that compared with heavy Higgs boson scatterings, axion productions from these SM-like Higgs boson scattering processes should be in the same order or less. 
Thus, these productions are not larger than the axion production from the heavy Higgs bosons and subdominant contributions. 

}

\subsection{Axion productions in the broken phase} \label{sec:Axion_productions_in_the_broken_phase}
\subsubsection{Productions from the SM-like Higgs boson decays}
\begin{figure}[t]
  \centering
  \includegraphics[scale=0.7]{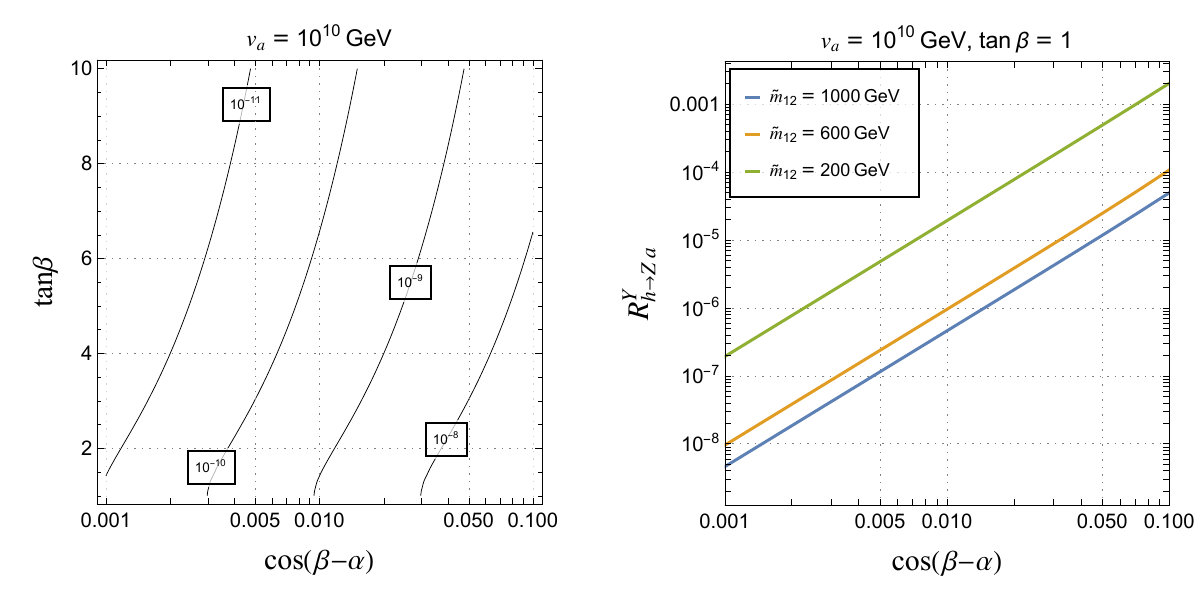}
  \caption{(Left panel): Contours of the yield $Y_a^{h\to a Z}$ from the SM-like Higgs boson decay $h\to Za$. 
  (Right panel): The ratio $R^Y_{h\to Z a}$ for $h\to Za$ as a function of $c_{\beta-\alpha}$. }
  \label{fig:yh}
\end{figure}

We here discuss axion thermal production from the SM-like Higgs boson decays.
As mentioned {at the beginning of Sec.~\ref{sec:axion productions}}, in the symmetric phase, SM-like Higgs boson decay into an axion is kinematically forbidden since the decay products necessarily contain heavy additional Higgs bosons.
In the broken phase, due to the mixing among CP-odd scalar states and the axion, the $hZa$ vertex arises. 
Thus, we calculate the Higgs boson decay $h\to Z a$ in the broken phase. 

The decay rate is calculated as
\bal
\Gamma(h\to a Z)\simeq\frac{m_h^3}{16\pi v_a^2}c_{\beta-\alpha}^2 s_{2\beta}^2
\left(1-\frac{m_Z^2}{m_h^2}\right)^3\;.
\eal
Similar to the heavy Higgs boson decay $h'_2\to a z_1'$, the yield for $h\to aZ$ is calculated by
\bal
Y^{h\to aZ}_a\simeq\frac{45 m_{\rm pl}}{1.66\cdot 4\pi^4}\frac{\Gamma_{h\to Za}}{m_h^2 } \int^{{x_{h, {\rm max.}}}}_{{x_{h, {\rm min.}}}}\frac{x_h^3}{g_\ast(x_h)^{3/2}}K_1\left(x_h\right)d x_h \;,
\eal
where $x$ is defined by ${x_h}=m_h/T$. 
We here set the lower limit of the integral by ${x_{h,{\rm min.}}}=m_h/(100{\rm GeV})$ since in the symmetric phase, this decay does not happen. 
The upper limit ${x_{h,{\rm max.}}} $  is taken to be infinity. 

The yield parameter $Y_a^{h\to aZ}$ is proportional to $c_{\beta-\alpha}^2$, so that it vanishes in the alignment limit $s_{\beta-\alpha}=1$. 
Also, the yield $Y_a^{h\to aZ}$ is suppressed by $c_\beta^2$ in the high $\tan\beta$ region. 
The numerical behavior of $Y_a^{h\to aZ}$ is displayed in the left panel of Fig.~\ref{fig:yh}, where we fix ${v_a}=10^{10}{\rm GeV}$ and varies $\tan\beta$ and $s_{\beta-\alpha}$.  

We also compare the size of $Y_a^{h\to aZ}$ with the yield from the heavy Higgs boson decays. 
The numerical result of the ratio $R^{Y}_{h\to Z a}$ is shown as a function of $c_{\beta-\alpha}$ in the right panel of Fig.~\ref{fig:yh}. We here set ${\tilde{m}_{12}}=m_H=m_{H^\pm}=200,\;600,\;$ and $1000{\rm GeV}$. 
Another parameter is fixed by ${v_a}=10^{10}{\rm GeV}$, and $t_\beta=1$. 
the ratio is maximized at $(c_{\beta-\alpha},{\tilde{m}_{12}})=(0.1,200{\rm GeV})$, it reaches $2\times 10^{-3}$. 
In this way, one can see that, even in the case of $s_{\beta-\alpha}\neq 1$, the axion production from SM-like Higgs boson is a subdominant process. 
\subsubsection{Productions through the top Yukawa couplings}
\begin{figure}[t]
  \centering
  \includegraphics[scale=0.7]{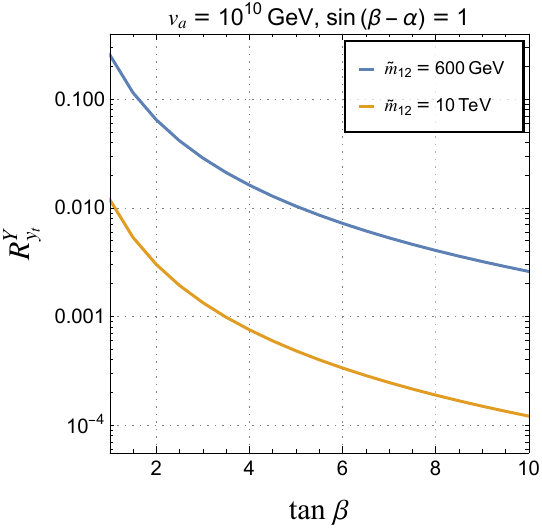}
  \caption{The ratio for the axion production through the top Yukawa coupling.  }
  \label{fig:ytop}
\end{figure}

While we focus on axion production from heavy Higgs boson decays and scatterings in the symmetric phase, axions can also be produced through scatterings among SM particles in the thermal plasma during the broken phase.
While we have calculated the $a$ production via the SM-like Higgs decay in the previous section, it vanishes in the alignment limit. 
To examine the axion productions in the broken phase with the alignment limit, we discuss the axion productions through the top Yukawa coupling. 
Following~Ref.~\cite{Salvio:2013iaa}, we here first describe the analytical expressions for the production rate for such a process. 
We then compare the amount of produced axion from these processes with those from heavy Higgs boson decays. 

The thermal production rate arising from the top Yukawa coupling is given by~\cite{Salvio:2013iaa},
\bal
\frac{Y_a^{y_t}}{ Y_a^{\rm eq} } =(1+r_{t}^{-3/2})^{-2/3} \simeq
 \left\{ \begin{array}{ll}
r_t &\hbox{for $r_t\ll1$}\\
1 & \hbox{for $r_t\gg 1$}
\end{array}\right.\label{eq:rrr}
\eal
where the yield in the thermal equilibrium $Y^{\rm eq}_a$ is given by $Y^{\rm eq}_a=n_a^{\rm eq}/S$ with $n^{\rm eq}_a=\zeta(3)T^3/\pi^2$ and the entropy density of the SM particles $S=2T^3 g_{\rm SM} \pi^2/45$. We set  $g_{\rm SM}=427/4$. 
The ratio $r_t$ is given in terms of production rates of axion $\gamma_a$, Hubble parameter $H$, the entropy density $S$ and $Y^{\rm eq}_a$  by
\bal
\label{eq:r}
r_t= \frac{2.4}{Y_a^{\rm eq} }\left.
\frac{\gamma_a}{Hs}\right|_{T=T_{\rm EW}} = 1.7 c_r^2 \frac{T_{\rm EW}}{10^{7}\GeV} \left(\frac{10^{11}\GeV}{v_a}\right)^2 c_t^{\prime 2}
\left. \frac{\gamma_a}{T^6 \zeta(3)/(2\pi)^5 (v_a/c_r)^2 c_t^{\prime 2}} \right|_{T=T_{\rm EW}}\; .
\eal
We set the temperature ($T_{\rm EW}$) at which the EWSB happens at 100GeV in the following numerical calculations. 
The coefficient is given by $c_r=2N~(1)$ for Type-II (I).
The axion coupling $c'_t$ is defined as the effective Lagrangian after performing a phase redefinition of fermion fields and Higgs fields (see the details in Ref.~\cite{Salvio:2013iaa}). 
For Type-I and Type-II, $c'_t$ is given by {$c'_t=X_{\phi_u}/c_r$}. 
The last factor in Eq.~\eqref{eq:r} is taken to be 71.8 at $T=100{\rm GeV}$, which is extrapolated from Fig.~5 of Ref.~\cite{Salvio:2013iaa}.

In contrast to the production from the SM-like Higgs decay, the production through the top Yukawa coupling is not suppressed by $c_{\beta-\alpha}$. 
The amount of the produced axion is governed by $\cos\beta$. 
To compare the yield $Y_a^{y_t}$ with the production from the heavy Higgs boson decays $Y_a^D$, we numerically calculate the ratio $R_{y_t}^Y\equiv Y_Y^{y_t}/Y_a^D$. The result {for Type-II} is shown as a function of $\tan\beta$ in Fig.~\ref{fig:ytop}\footnote{{We checked that the result of Type-I is almost the same as Type-II.}}, where we fix the input parameter ${v_a}=10^{10}{\rm GeV}, s_{\beta-\alpha}=1$, $\tilde{m}_{12}=m_{H}=m_{H^\pm}=600{\rm GeV}$ and $10{\rm TeV}$. 
As seen in the figure, the ratio $R^Y_{y_t}$ reaches $\sim0.25$ at $(t_\beta,{\tilde{m}_{12}})=(1,600{\rm GeV})$, which is larger than the dominant contributions from the heavy Higgs boson scattering, i.e., $H'_2V\to H'_1a$ ($R_{H'_2V\to H'_1a}^Y\sim 0.12$ at ${\tilde{m}_{12}}=600{\rm GeV}$ in Fig.~\ref{FIG:ratio of scattering and decay}). 
Thus, the production through the top Yukawa coupling is comparable with the productions from heavy Higgs boson scattering. 
However, it does not exceed the production from the heavy Higgs boson decays under the circumstances that the reheating temperature is larger than the masses of the heavy Higgs bosons.

\section{{Axion contribution to {$\Delta N_{\rm eff}$}}}\label{sec:delta_Neff}

\begin{figure}[tb]
  \centering
  \includegraphics[scale=0.7]{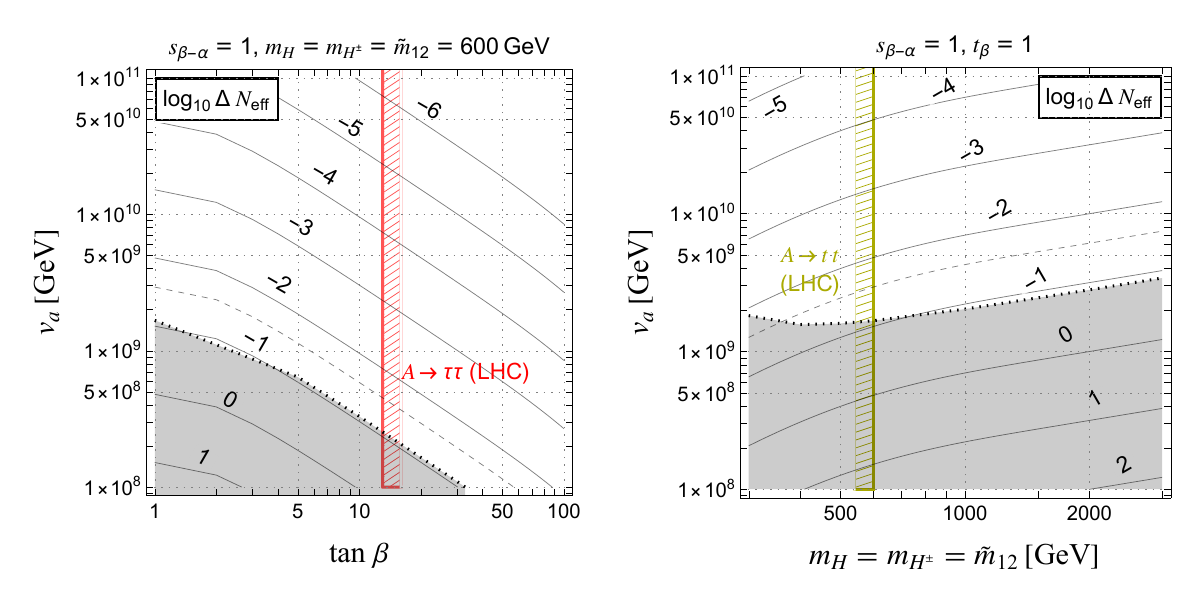}
  \caption{
The effective number of neutrino spices subtracted by the SM neutrinos $\Delta N_{\rm eff}$ in the plane of $(v_a, \tan\beta)$ and $(v_a, m_H)$. 
The red and yellow lines show the constraint of the direct searches of the heavy Higgs bosons $A\to \tau\tau$ and $A\to tt$ for Type-II, respectively~\cite{Aiko:2020ksl}. 
Below the dashed line, the predicted axion abundance \eqref{eq:Yh2toh1a} exceeds the equilibrium value, $\Delta N_{\rm eff} \gtrsim 0.03$.
In the gray-shaded region, 
the axion production rate via scattering at $T = m_H/2$ becomes greater than the Hubble parameter, and the axions are considered to be in equilibrium, $\Delta N_{\rm eff} \simeq 0.03$.
} 
  \label{fig:DNeff}
\end{figure}

In this section, we discuss the dark radiation of axions produced from heavy Higgs boson decays.
As we have already discussed, the axion production from scattering processes is a subdominant contribution, so we neglect it here for simplicity.

Using the axion yield, one can also compute the effective number of neutrino species beyond the SM neutrinos, which is defined by
\bal
\label{eq:DNeff}
\left.\Delta N_{\rm eff}\equiv \left( \frac{\rho_a}{\rho^{(1)}_\nu} \right) \right|_{\rm MeV} \;. 
\eal
We estimate {$\rho_a|_{\rm MeV}$} by
\bal\label{eq:rho_a}
{\left.\frac{\rho_a}{S}\right|_{\rm MeV}} \sim E_{a}[M_{H'_2}]
\left(\frac{a_s[\rm MeV]}{a_s[ {M_{H'_2}}]} \right)^{-1}Y^D_a \;,
\eal
where $a_s[{\rm MeV}]$ denotes the scale factor at $T=1{\rm MeV}$ and $a_s[M_{H_2'}]$ is the one at $T=M_{H_2'}$. 
{From {the  conservation of entropy}, the scaling factor can be reduced as $(a[{\rm MeV}]/a[M_{H_2}'])^{-1}=(g_{\ast S}[{\rm MeV}]/g_{\ast S}[M_{H_2}'])^{1/3}(T_{\rm MeV}/M_{H_2}')$ with $T_{\rm MeV}=1{\rm MeV}.$}
The energy density of {a single neutrino species} $\rho_\nu^{(1)}$ is given by {$\rho_\nu=(7/8)2(\pi^2/30)T^4$}. 
Since the axion just after the creation is relativistic, we set $E_{a}[M_{H'_2}]\sim  M_{H_2'}${/2}.
Substituting Eq.~\eqref{eq:rho_a} into Eq.~\eqref{eq:DNeff}, one obtains 
\bal\label{eq:DNeff reduced}
\Delta N_{\rm eff}\simeq 
{\frac{8}{21}}
\frac{g_{\ast S}^{4/3}[{\rm MeV}]}{g_{\ast S}^{1/3}[M_{H'_2}]}Y^D_a\;,
\eal
where we set $g_{\ast S}[{\rm MeV}]=10.75$, $g_{\ast S}[M_{H'_2}]=106.75$. 
The quantity $\Delta N_{\rm eff}$ counts the degrees of freedom of additional relativistic species beyond the SM.
We note that $\Delta N_{\rm eff}$ can be estimated by Eq.~\eqref{eq:DNeff reduced} for all axions with masses {$m_a\ll 1{\rm MeV}$ to apply constraints from BBN}.
{To apply constraints from CMB, one should evaluate $\Delta N_{\rm eff}$ at $T\sim 0.1{\rm eV}$ and consider axions with $m_a\ll 0.1{\rm eV}$.}
Since $\Delta N_{\rm eff}$ scales with $Y_a^D$, the model parameter dependence of $\Delta N_{\rm eff}$ is the same as that of $Y_a^D$.
Note that we estimate the high-energy contribution to $\Delta N_{\rm eff}$. In fact, axions can also be produced around the QCD crossover through scattering with pions~\cite{DEramo:2021lgb}, which is not considered here.

We present the numerical results of $\Delta N_{\rm eff}$ for Type-II in Fig.~\ref{fig:DNeff}, where we set $s_{\beta-\alpha}=1$ and consider a degenerate mass scenario for the heavy Higgs bosons, with $m_{H}=m_{H^\pm}=\tilde{m}_{12}$.
In each plot, two of the three parameters $v_a$, $t_{\beta}$, and $m_H$ are scanned.
Variations in $\Delta N_{\rm eff}$ due to changes in $s_{\beta-\alpha}$ are small.
We have verified that similar numerical results are obtained for Type-I.
While we {focus on} the scenario that the axion is not fully thermalized, $\Delta N_{\rm eff}$ is  $(4/7)(g_{\ast S}[{\rm MeV}]/g_{\ast S}[M_{H_2^\prime}])^{4/3}\simeq0.03$ under the assumption that the axion is in thermal equilibrium {at temperatures above the weak scale}. 
Hence, in the region with $\Delta N_{\rm eff}\gtrsim 0.03$, our computation of $\Delta N_{\rm eff}$  may not be {simply} applicable. 
In such a parameter region, the effect of back reaction {through scattering processes is considered} to become significant {especially for small $v_a$. 
}
{These effects are} not included in Eq.\eqref{eq:Yh2toh1a}.
To improve the evaluation of $Y^{(D)}$, one should incorporate {these} effects, which is beyond the scope of this paper. {
In the region below the dashed line in Fig.~\ref{fig:DNeff},  the predicted axion abundance exceeds $\Delta N_{\rm eff}\simeq 0.03$. In the gray-shaded region, the axion production rate through scattering at $T = m_H/2$ exceeds the Hubble parameter, and so, we expect $\Delta N_{\rm eff}\simeq 0.03$ there.
Thus, below the dashed line, 
the axion abundance is considered to be comparable to or larger than the thermal equilibrium value.
}

In the figure, we display the bounds from direct searches for heavy Higgs bosons at LHC Run I for Type-II~\cite{Aiko:2020ksl}.
The $A\to\tau\tau$ channel (red line) provides an upper bound for $\tan\beta$,
while the $A\to tt$ channel (yellow line) gives a lower bound for the heavy Higgs boson mass.
Interestingly, the parameter region with a larger $\Delta N_{\rm eff}$ tends to lie far from the exclusion regions by direct searches of heavy Higgs bosons. 
{Thus, future CMB experiments such as CMB-S4~\cite{CMB-S4:2016ple, Abazajian:2019eic} {could} give a complementary constraint to those from the direct searches of heavy Higgs. More specifically, it could constrain parameter regions that are challenging to probe through the collider searches, i.e., low $\tan\beta$ region and large mass regions of heavy Higgs.}
{To summarize this section}, axions with sufficiently small mass to affect $\Delta N_{\rm eff}$ contribute a relativistic degree of freedom as dark radiation when produced from heavy Higgs bosons.
We emphasize that this discussion is general and can be applied to the canonical DFSZ axion model.

\section{Cosmological constraints for axion with a mass of keV to sub-GeV scale} \label{sec:cosmological constraints}
Axion production predominantly occurs through heavy Higgs boson decays rather than scattering processes under {the assumptions described at the beginning of Sec.~\ref{sec:axion productions}}. In this section, we focus on axions with masses in the keV to sub-GeV range. This mass range is particularly interesting because it is subject to various cosmological constraints. Moreover, since the axion production mechanism we consider is universal for DFSZ-type axions, we can impose highly general constraints on this class of axions.

 For $m_a < 2m_e$, axions decay primarily into two photons, while decays into electrons and positrons become kinematically allowed for higher masses.
Axions with keV-scale masses produce photons upon decay, which can be detected by X-ray searches. This provides a powerful probe for axions in this mass range. For MeV to sub-GeV scale axions, the decay products (photons, electrons, and positrons) can be constrained by observations of the CMB and by requiring consistency with BBN.
These cosmological constraints are particularly valuable for DFSZ-type axions because (i) the production mechanism we have discussed is universal for DFSZ-type axions, occurring through their coupling to heavy Higgs bosons; (ii)
the constraints apply over a wide range of model parameters, making them highly general.

In the following subsections, we will discuss each of these constraints in detail, examining their implications for DFSZ-type axion models and their interplay with the production mechanisms we have studied.

\subsection{Constraint from X-rays}

\begin{figure}[t]
  \centering
  \includegraphics[scale=0.65]{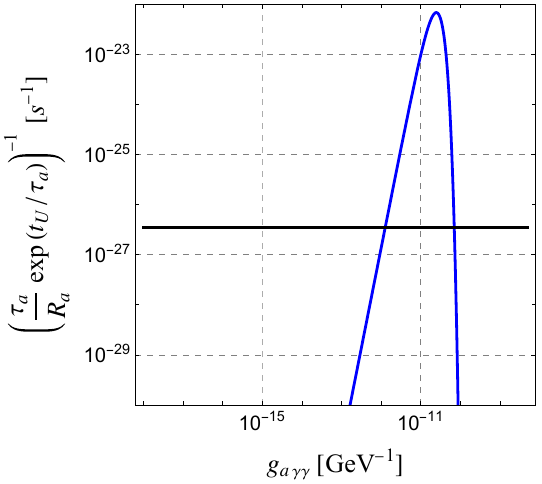}\hspace{2mm}
  \includegraphics[scale=0.65]{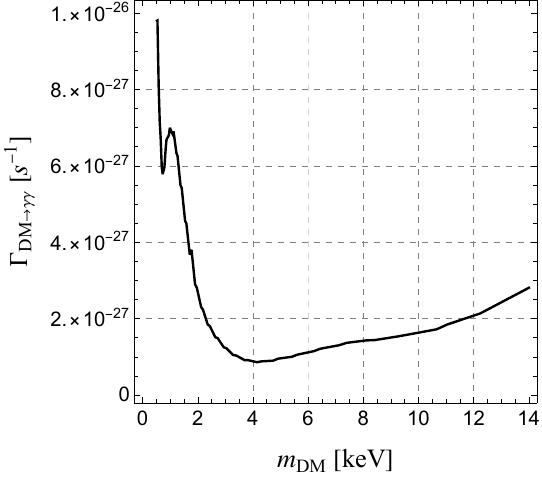}
  \caption{
    (Left panel): The behavior of the inverse of the lifetime $\tau_a$ multiplied the factor ${\cal R}_a \exp(\tau_a/t_U)$ is shown by the blue line. 
    We choose $m_a=$1 keV, $s_{\beta-\alpha}$=1 $t_{\beta}=5$,$m_H=\tilde{m}_{12}=500{\rm GeV}$. 
    The black line shows the upper bound on the $\Gamma_{\rm DM \to \gamma \gamma}$ (see the right panel). 
    (right panel): The lower bound on the decay rate $\Gamma_{\rm DM \to \gamma \gamma}$ for the decaying DM derived from the observations of Large Magellanic Clouds (LMC) from the XMM-Newton~\cite{Boyarsky:2006ag}.  
   }
  \label{FIG:LMC}
 \end{figure}
 
The axion with a mass of keV scale is constrained by X-ray observations.
We here discuss constraints for the axion coupling and axion mass from X-ray observations following Refs.~\cite{Takahashi:2020bpq,Langhoff:2022bij}. 
We do not require the axion constitute of all cold dark matter (CDM), i.e., $\Omega_{a} \leq \Omega_{\rm CDM}^{obs.}$.

The axion density at a time $t$ is written by 
\bal
\rho_a=e^{-\frac{t-t_U}{\tau_a}} \rho_{a,0} {a_s^{-3}},
\eal
where ${a_s}$ denotes the scale factor normalized to be unity at present, $\rho_{a,0}$ denotes the current density of axion, and $t_U$ is the current age of the Universe. The energy density is related to the yield parameter as {$\rho_{a,0}\simeq S_0 m_a Y_a$} where {$S_0$} is the entropy at present. 
The exponential factor takes into account depletion by the decay of axion. 
Introducing the density of DM  $\rho_{\rm DM}$, the above expression is further reduced to
\bal
\rho_a={{\cal R}_a}e^{-\frac{t}{\tau_a}} {\rho_{{\rm DM}}},
\eal
here we have introduced the fraction parameter for the axion density {${\cal R}_a$}, 
{which quantifies the fraction of the axion density before the decays:}
\bal \label{eq:Ra definiton}
{{\cal R}_a}=\frac{\rho^{\tau_a\to \infty}_{a,0}}{\rho_{{\rm DM},0}}\;. 
\eal
 One can express $\rho_{a,0}$ by using the axion density before the decay $\rho^{\tau\to \infty}_{a,0}$ as
\bal\label{eq:rho_a_to_rho_a_0}
\rho_{a,0}=e^{-\frac{t_U}{\tau_a}} \rho^{\tau_a\to \infty}_{{ a},0}\;.
\eal
Using this expression,  we can see that ${\cal R}_a$  is {given by}
\bal \label{eq:fraction_energy_density}
{{\cal R}_a}=e^{t_U/\tau_a}\frac{\rho_{a,0}}{\rho_{{\rm DM},0}}\;. 
\eal

Let us formulate the constraint from the X-ray for the decaying axion. 
The starting point is the differential photon flux produced from the DM decay,~\cite{Lisanti:2017qoz} 
\bal\label{eq:gamflux}
\frac{d\Phi}{dE}
=\left(\frac{1}{4\pi m_{\rm DM}\tau_{\rm DM}}\sum_i {\rm BR}_i\frac{dN_i}{dE}\right)\times D
\eal
where ${\rm BR}_i$ is the branching fraction for DM decays and ${dN_i}/{dE}$ denotes the corresponding photon energy distributions. 
The $D$ factor is defined by
\bal
D=\int ds \,d\Omega \,\rho_{\rm DM}(s,\Omega). 
\eal
The unit of the flux is 
$[{d\Phi}/{dE}]=
{\rm count}\cdot s^{-1}\cdot {\rm GeV}^{-2}\cdot {\rm sr}^{-1}$. 
The flux for the decaying axion can be derived by the replacements $m_{\rm DM}\to m_a$, $\tau_{\rm DM}\to \tau_{a\to\gamma\gamma}$, $\rho_{\rm DM}\to \rho_{a}=e^{-t/\tau_a}{{\cal R}_a}\rho_{\rm DM}$ in Eq.~\eqref{eq:gamflux}. 
Therefore, for the observed lower limit for the lifetime of DM from X-ray observations, $\tau_{\rm DM}^{\rm min}$,
the following inequality should be satisfied,
\bal\label{eq:ineqxray}
\tau_{\rm DM}^{\rm min}<\tau_{a\to\gamma\gamma}\frac{\exp \left({{t_U}}/{\tau_a}\right)}{{\cal R}_a}. 
\eal  
The right-handed side is governed by the fraction parameter {${\cal R}_a$} and the lifetime of axion $\tau_a$.
From physical intuition, one may expect that the photon flux produced from the decaying axion is reduced if the amount of axion in the Universe is small, leading to a weaker X-ray constraint.
This intuition coincides with the scaling of {${\cal R}_a$} in Eq.~\eqref{eq:ineqxray}. 
{
Another case where the limit of X-rays is weaker is the case of relatively smaller $\tau_a$ than the age of the Universe. 
In this case, even if the fraction  {${\cal R}_a$} is sufficiently large, the influence of the axion decay into di-photon on the X-ray observations becomes mild. 
This is because the axion abundance at the present time is small due to the depletion of the decay, as can be seen in Eq.~\eqref{eq:rho_a_to_rho_a_0}. 
}
We will present the explored region of our model by the X-ray observations in the next section (Fig.~\ref{FIG:2HDMsC}). 
In this analysis, we make use of the experimental constraints for $\tau_{\rm DM}^{\rm min}$ derived by XMM-Newton~\cite{Boyarsky:2006ag,Boyarsky:2006fg,Foster:2021ngm}, NusSTAR~\cite{Roach:2022lgo,Roach:2019ctw,Ng:2019gch}, and Integral~\cite{Laha:2020ivk} observations. 
These experiments observed the Milky Way, M31, and the Large Magellanic Clouds (LMC). 

Before we close this section, let us demonstrate how one can obtain the bound on the axion-photon coupling from the inequality Eq.~\eqref{eq:ineqxray}. 
To do this, we here fix the model parameter as $m_a=$1 keV, $s_{\beta-\alpha}$=1, $t_{\beta}=5$, and $m_H={\tilde{m}_{12}}=500{\rm GeV}$. 
For this axion, the constraint from XMM-Newton~\cite{Boyarsky:2006ag} is relevant. 
The upper bound on the decay rate for DM decaying two photons is given in the right panel of Fig.~\ref{FIG:LMC}. 
In the left plot, the behavior of the inverse of the right-handed side of Eq.~\eqref{eq:ineqxray} is shown by the blue line. 
 As seen, it maximized at $g_{a\gamma \gamma}^{\rm max}\simeq 5\times 10^{-11}{{\rm GeV}^{-1}}$. 
 In the region $g_{a\gamma \gamma}\lesssim g_{a\gamma \gamma}^{\rm max}$, $\tau_a \gtrsim t_U$, so that 
 the exponential factor is close to unity and the factor $(\tau_a/{\cal R}_a)^{-1}$ decreases as $g_{a\gamma \gamma}$ becomes small. 
 If  $g_{a\gamma \gamma}> g_{a\gamma \gamma}^{\rm max}$, $\tau_a \lesssim t_U$ and the suppression by $\exp(t_U/\tau_a)^{-1}$ is significant. 
 The black horizontal line shows the upper bound on $\Gamma_{\rm DM \to \gamma \gamma}$ for $m_{\rm DM}=1$ keV. 
The above this horizontal line, i.e., $1\times 10^{-12}{{\rm GeV}^{-1}}\lesssim g_{a\gamma \gamma}\lesssim9\times 10^{-11}{{\rm GeV}^{-1}}$ is excluded by the the observation by XMM-Neuton.  
Although the axion with $m_a=1 {\rm keV}$ is discussed here, one can obtain the explored region by the X-ray observations in the plane of ($m_a, g_{a\gamma\gamma}$) by applying this procedure to another value of $m_a$. 

\subsection{CMB constraints}
\subsubsection{Spectral distortion}
It is known that the spectrum of CMB is very close to the blackbody radiation with ${T_{\rm CMB}}=2.725 {\rm K}$. 
The departure of the CMB spectrum from the blackbody radiation is called spectral distortions (SD), which can probe the thermal history of the Universe during pre- and post-recombination (reviews can be found in Refs.~\cite{Tashiro:2014pga,Chluba:2018cww}. See also the recent pedagogical introduction~\cite{Lucca:2019rxf}).
We apply the bound of SDs in the epoch of the red shift of $10^3 \lesssim z \lesssim 10^6$. 
The upper bound of the redshift corresponds to the era when the so-called $\mu$ distribution arises. The lower bound corresponds to the recombination era. 
No discovery of the SDs gives constraints to the model parameter space. 
Since axion can decay into two photons during the above epoch of the redshift, it can contribute to the SDs in case the lifetime is in the range of  
\bal\label{eq:SDr}
2.4\times 10^{7}s\lesssim \tau_a \lesssim {1.4}\times 10^{13}s.
\eal
Here, the lower bound has been derived by using the Hubble parameter in the radiation-dominant era
\bal
H=\frac{1}{2t}=\frac{\pi \sqrt{8\pi g_{\ast}}}{\sqrt{90}}\frac{T^2_{\rm CMB}}{m_{\rm pl}}(1+z)^2\;,
\eal
where we set the effective degree of freedom $g_{\ast}=3.363$ and the temperature of the CMB $T_{\rm CMB}={2.725}$K, the Planck mass $m_{\rm pl}=1.22\times 10^{19}$ GeV, and $z= 10^{6}$. 
For the upper bound, we have set $z=10^3$ and applied the Hubble parameter under the assumption of $\Omega_{r0},\Omega_{m0}a \gg \Omega_{K0}a^2,\Omega_{\Lambda}a^4$, i.e.,
\bal
H&=H_0\frac{\Omega_{r0}^{1/2}+ \Omega_{m0}\theta}{(2\Omega_{r0}^{1/2}\theta+\Omega_{m0}\theta^2)^2}
=\sqrt{\frac{4\pi^3{g_\ast}}{45}}\frac{T_{\rm CMB}^2}{m_{\rm pl}}(1+z)^{3/2}(2+z+z_{\rm eq})^{1/2}
\;, \notag \\
H_0t&=2 \Omega^{1/2}_{r0}\theta^2+\frac{2}{3}\Omega_{m0}\theta^3
,\eal
with $\theta$ being the parametric variable and $z_{\rm eq}=\Omega_{m0}/\Omega_{r0}-1$ being the redshift at the matter-radiation equality time, 
where the density parameters $\Omega_{r0}$ and $\Omega_{m0}$, and the Hubble parameter at the current time are taken from Ref.~\cite{Planck:2018vyg}.
In Fig.~5 of Ref.~\cite{Poulin:2016anj}, the upper bounds of the effective energy density $\Xi$ (the definition is in the reference) normalized by the CDM energy density from the SDs are calculated for given a lifetime of DM. 
In our case $\Xi={\cal R}_a$ holds and the fraction parameter {${\cal R}_a$} and the lifetime $\tau_a$ can be regarded as the function of $g_{a\gamma\gamma}$ and $m_a$. 
Thus, we can recast this bound into the upper bound of $g_{a\gamma\gamma}$ for a given $m_a$. 
The result will be given in Fig.~\ref{FIG:2HDMsC}.

\subsubsection{CMB anisotropies}
\begin{figure}[t]
  \centering
  \includegraphics[scale=0.85]{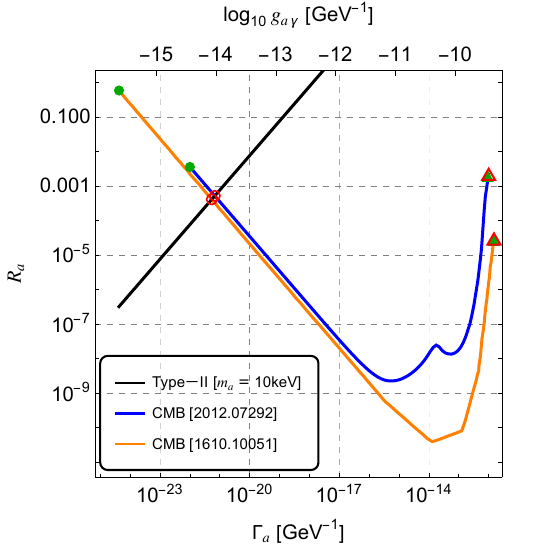}
  \caption{
    Constraint for the fraction of the axion relic density from CMB anisotropy.
    The upper limit from Ref.~\cite{Bolliet:2020ofj} (Ref.~\cite{Poulin:2016anj}) is shown by the blue (orange) line. 
    Black line denotes the theory calculations in Type-II with $m_a=10{\rm keV},~t_{\beta}=1,~m_{H}=m_{A}=m_{H^\pm}=500{\rm GeV}$. 
    The red circle ($\circ$) and triangle (\scalebox{0.7}{$\triangle$}) show the lower and upper bounds for $\Gamma_{a}$ in Type-II. 
    Similarly, the green circle ($\bullet$) and triangle (\scalebox{0.7}{$\blacktriangle$}) show the lower and upper bounds for $\Gamma_{a}$ in Type-I.
     }
  \label{FIG:CMBanti}
 \end{figure}
Axion can also be constrained by CMB anisotropy since the decay of the axion can affect the ionization history around the last scattering. 
The bound is relevant for longer lifetimes of the axion than those given in Eq.~\eqref{eq:SDr}.  
In order to take into account the constraint from CMB anisotropies, we recast the exclusion limits in the plane of $({\cal R}_a, \tau_a)$ obtained in Refs~\cite{Bolliet:2020ofj} and \cite{Poulin:2016anj}. 
The former (latter) is relevant for the mass range $m_a\lesssim20{\rm keV}$ ($10 {\rm keV} \lesssim m_a\lesssim1 {\rm TeV}$). 
In Fig.~\ref{FIG:CMBanti}, we show the upper bounds for the fraction of the axion relic density {${\cal R}_a$} analyzed in the above-mentioned references. 
We here take the input parameters as $m_a=10{\rm keV},~t_{\beta}=1,~m_{H}=m_{A}=m_{H^\pm}=500{\rm GeV}$. 
The model prediction of Type-II is shown by the black line (For simplicity of the computation, we here only include the contributions from the heavy Higgs decays). 
For this benchmark point, the limit $7\times 10^{-22}{{\rm GeV}^{-1}}\lesssim\Gamma_a^{\rm TypeII}\lesssim 2\times 10^{-12}{{\rm GeV}^{-1}}$ is extracted as shown by the red circles and triangles. 
This is translated into the limit for $g_{a\gamma\gamma}$, $1\times 10^{-14}{{\rm GeV}^{-1}}\lesssim g_{a\gamma\gamma}\lesssim 4\times 10^{-10}{{\rm GeV}^{-1}}$. 
For Type-I with this benchmark point, the model predictions of {${\cal R}_a$} are much greater than the upper limits of the CMB anisotropy. 
This can be understood from the scaling of ${\cal R}_a^{\rm TypeI}$
\bal
 {\cal R}_a^{\rm TypeI}\simeq \left(\frac{E_{\rm TypeII}}{E^\prime_{\rm TypeI}}\right)^2 {\cal R}_a^{\rm TypeII}\simeq6.3\times 10^{10}\left(\frac{10{\rm keV}}{m_a}\right)^4\left(\frac{1/\sqrt{2}}{\cos\beta}\right)^4{\cal R}_a^{\rm TypeII},
\eal
which can be extracted from Eqs.~\eqref{eq:agg_agamgam}, \eqref{eq:E_tyII} and \eqref{eq:Ep_tyI}. 
Hence, the limit of the decay rate is given by endpoints of observation of the CMB anisotropy denoted by the green circles and triangles in the figure, i.e., 
$4\times 10^{-25}{{\rm GeV}^{-1}}\lesssim\Gamma_a^{\rm TypeI}\lesssim 2\times 10^{-12}{{\rm GeV}^{-1}}$. 
While we have discussed the bound for a fixed mass of the axion, the excluded region of the CMB anisotropy {in varying $m_a$} will be shown in Fig.~\ref{FIG:2HDMsC}.

\subsection{BBN}\label{sec:BBN}

The constraint from BBN for the axion is discussed in Ref.~\cite{Depta:2020wmr}. 
In this reference, it is assumed that the axion interaction with the photon is dominant. 
Two most relevant processes for the axion productions, Primakoff process $q^\pm a \leftrightarrow q^\pm \gamma$ and inverse decays, $q^\pm q^\pm\to a$, are taken into account, where $q^{\pm}$ denotes the charged particles in thermal plasma.
We make use of the results of $T_{\rm RH}= 100~{\rm GeV}$ in Fig.~5 of Ref.~\cite{Depta:2020wmr} to apply the BBN bound. 
As shown in the plot, at least the lifetime of $\tau_a<10^9$s is excluded. 
Following this, we apply the upper bound of the lifetime of the axion by $\tau_a>10^9$s as the conservative limit. 
Since the lifetime is a function of $m_a$ and $g_{a\gamma\gamma}$, this limit of the lifetime can be mapped into the plane of $(m_a, g_{a\gamma\gamma})$. 
We note that in our scenario, there are thermal productions from the heavy Higgs bosons in addition to the Primakoff process and the inverse decays of the SM particles. 
If one adds it to the analysis of BBN constraint, the limit would become more tighter. 
The excluded region from the BBN in the plane of $g_{a\gamma\gamma}$ and $m_a$ is shown by the green region in Fig.~\ref{FIG:2HDMsC}.

\subsection{Numerical results} \label{sec:numerical results}

\begin{figure}[t]
  \centering
  \includegraphics[scale=0.75]{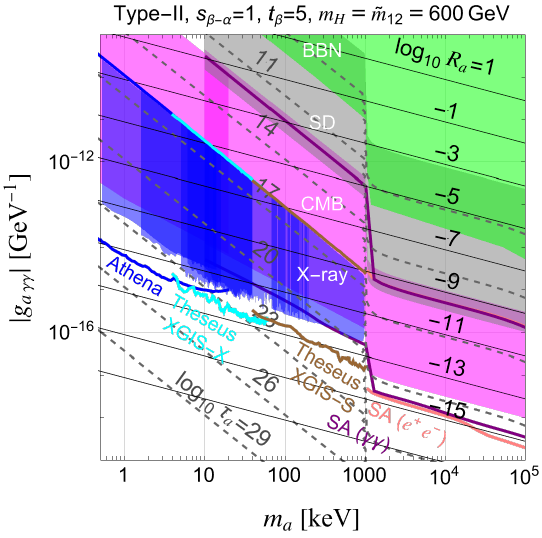}
    \includegraphics[scale=0.75]{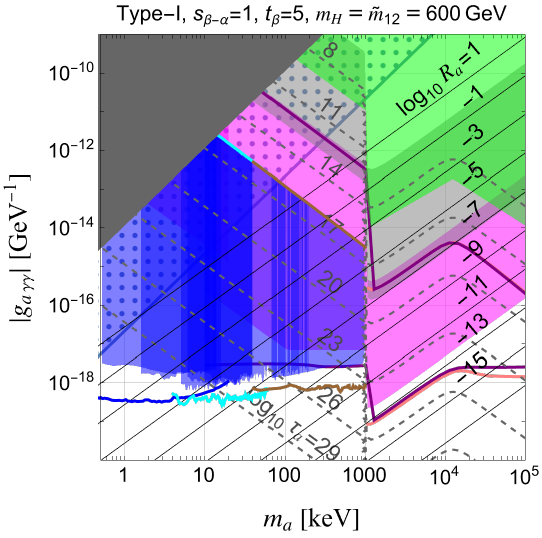}\\
      \includegraphics[scale=0.75]{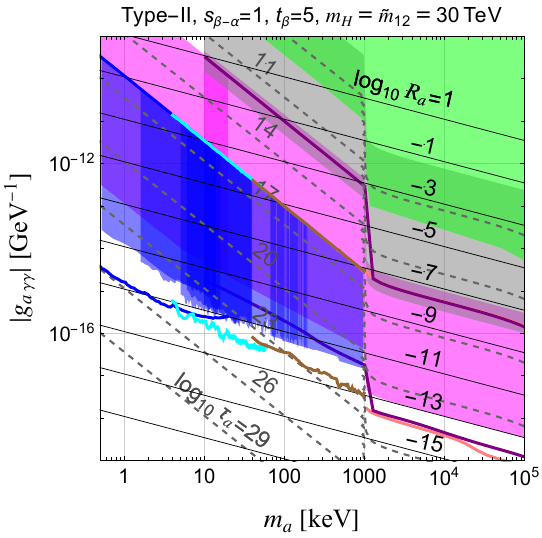}
    \includegraphics[scale=0.75]{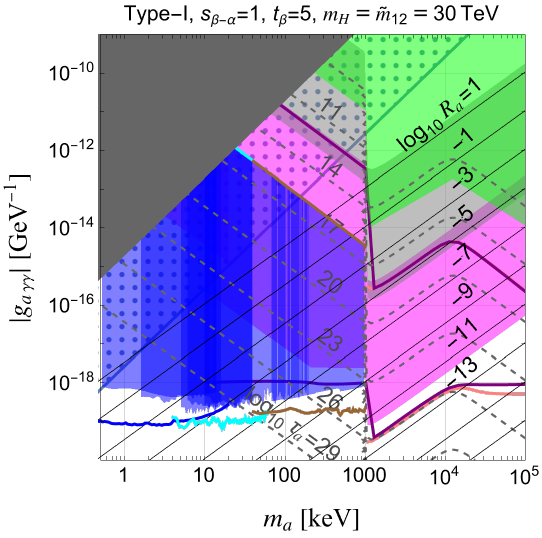}
  \caption{Constraints of various astrophysical observations for the model of Type-I and Type-II DFSZ-type axion models in the plane of $g_{a\gamma \gamma}$ and $m_a$, where benchmark scenarios A (Eq.~\eqref{eq:scA}) and B (Eq.~\eqref{eq:scB}) are chosen in the top panels and bottom panels, respectively. The colored shaded region shows the excluded regions by X-ray observations (blue), CMB anisotropies (pink), CMB spectral distortions (gray), and BBN (green). 
  Black solid lines show the fraction for the relic density {${\cal R}_a$} defined in Eq.~\eqref{eq:fraction_energy_density} (the black dashed lines are the lifetime of the axion). 
  For Type-I, the dark gray region indicates {the parameter space where the VEV of the complex singlet $v_s$ gets smaller than the heavy Higgs boson mass, i.e., $v_S<m_H={\tilde{m}_{12}}$. (In Sec.~\ref{sec:model}, we have assumed that $\rho$ is heavier than the additional Higgs bosons.)}
  The solid lines also show future sensitivities of X-ray and CMB anisotropies, such as Athena (blue), TESEUS XGIS-X (cyan) TESEUS XGIS-S (brown), and  $a\to e^+e^-$ and $a\to \gamma\gamma$   
 searches of Simons array (SA). 
{Dotted regions show the regions outside the applicable range for our computations of freeze-in axion production.} 
}
  \label{FIG:2HDMsC}
 \end{figure}

\begin{figure}[t]
  \centering
  \includegraphics[scale=0.75]{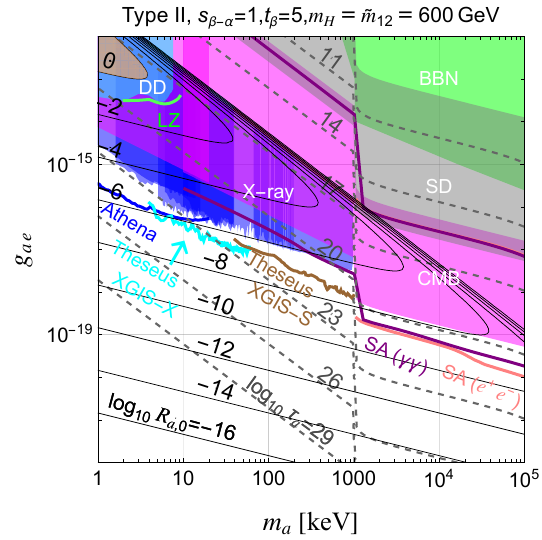}
    \includegraphics[scale=0.75]{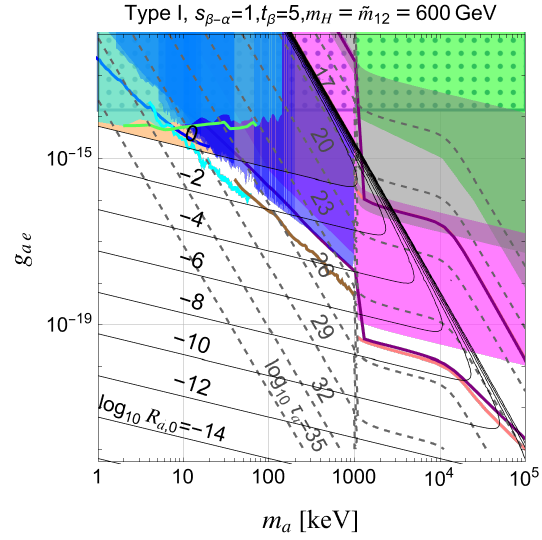}\\
      \includegraphics[scale=0.75]{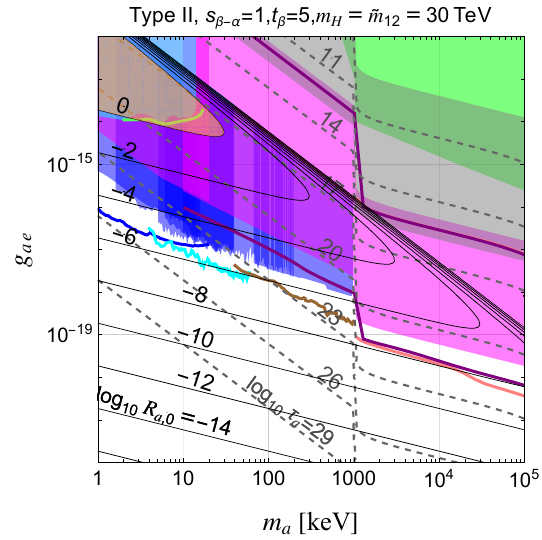}
    \includegraphics[scale=0.75]{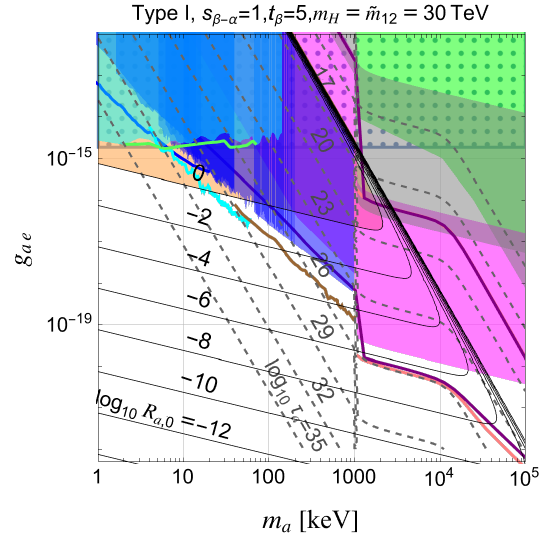}
  \caption{Constraints of various astrophysical observations and direct detection experiments for the model of Type-I and Type-II DFSZ-type axion models in the plane of $g_{ae}$ and $m_a$, where benchmark scenarios A (Eq.~\eqref{eq:scA}) and B (Eq.~\eqref{eq:scB}) are chosen in the top panels and bottom panels, respectively. The colored shaded region shows the excluded regions by X-ray observations (blue), CMB anisotropies (pink), CMB spectral distortions (gray), and BBN (green), and the direct detection (DD) experiment of XENONnT (cyan). 
  Black solid lines show the fraction for the relic density ${\cal R}_{a,0}$ defined in Eq.~\eqref{eq:def R_a 0} (the black dashed lines are the lifetime of the axion). 
  For Type-I, the orange region indicates the parameter space where the axion abundance is larger than the observed relic density of DM, i.e., $\rho_{a,0}>\rho_{{\rm DM},0}$.
  The solid lines also show future sensitivities of X-ray and CMB anisotropies, such as Athena (blue), TESEUS XGIS-X (cyan), TESEUS XGIS-S (brown), $a\to e^+e^-$ and $a\to \gamma\gamma$   
 searches of Simons array (SA) and LZ (green). 
{Dotted regions show the regions outside the applicable range for our computations of freeze-in axion production.}
}
  \label{FIG:2HDMsC_gaee}
 \end{figure}

Drawing upon the above discussions, we now examine the regions excluded by cosmological observations for keV to sub-GeV scale axions produced through heavy Higgs boson decays and scatterings.

To set the benchmark points for this analysis, let us first describe the experimental bound on the model parameter space for the Higgs sector. 
One of the most relevant constraints comes from the measurements of the 125 GeV Higgs boson in LHC
~\cite{Chen:2013kt,Celis:2013rcs,Chiang:2013ixa,Grinstein:2013npa,Eberhardt:2013uba,Chang:2013ona,Bernon:2015wef,Chowdhury:2015yja,Bernon:2015qea,Cacchio:2016qyh,Belusca-Maito:2016dqe}. 
Since current data of Higgs boson measurements indicates that the discovered Higgs boson is SM-like, the alignment limit $s_{\beta-\alpha}=1$ is favored. 
While the slight deviations from $s_{\beta-\alpha}=1$ can be consistent with the experimental data, we focus on the alignment limit. 
This is indeed a reasonable scenario for axion thermal productions in the sense that the axion relic abundance is maximized at $s_{\beta-\alpha}=1$ (see the left bottom plot of Fig.~\ref{fig:omegaD}). 
Another important collider constraint is direct searches of the additional Higgs bosons
~\cite{Kanemura:2014bqa,Craig:2015jba,Bernon:2015qea,Bernon:2015wef,Cacchio:2016qyh,Sanyal:2019xcp,Karmakar:2019vnq,Arco:2020ucn,Blasi:2017zel,Aiko:2020ksl}.
Following the results of Ref.~\cite{Aiko:2020ksl}, the relevant channels are $pp\to H^\pm \to tb$ and $pp\to A\to \tau\tau $ in the alignment limit. 
The former gives the lower bound $1\lesssim t_\beta$ at $m_H=600{\rm GeV}$ for all the types of the 2HDMs. 
The latter gives the upper bound $ t_\beta\lesssim 12$ for Type-II at $m_{H}=600{\rm GeV}$. 
One should also take into account flavor constraints~\cite{Misiak:2017bgg,Misiak:2020vlo,Cheng:2015yfu,Enomoto:2015wbn,Haller:2018nnx}. 
For Type-II, the stringent bounds for the charged Higgs boson mass, $m_{H^\pm}\lesssim 580 {\rm }$ GeV, is imposed from $B_s\to X_s\gamma$~\cite{Misiak:2017bgg}. 
For Type-I, the strongest bound comes from $B_{d}\to \mu\mu$, by which the lower bound of $t_\beta$ is imposed, e.g., $1.5 \lesssim t_\beta$ at $m_{H^\pm}=600$ GeV~\cite{Haller:2018nnx}. 
We also note that the large mass hierarchy among additional Higgs boson masses would conflict with the precision test of the electroweak parameters(e.g., see \cite{Grimus:2007if,Grimus:2008nb,Kanemura:2011sj}). 
Considering these experimental constraints for the Higgs sector, we focus on the two distinct scenarios
\bal
\label{eq:scA}
&{\rm (A)}~~ m_H=m_{H^\pm}={\tilde{m}_{12}}=600{\rm GeV}, 
\\  \label{eq:scB}
&{\rm (B)}~~ m_H=m_{H^\pm}={\tilde{m}_{12}}=30{\rm TeV}.
\eal
For these scenarios, Higgs mixing parameters are commonly fixed by $s_{\beta-\alpha}=1,\;\; t_{\beta}=5$. 
Scenario A corresponds to the relatively light mass spectrum of the additional Higgs boson satisfying the above-mentioned experimental constraints. 
Scenario B is the heavy scenario of the additional Higgs bosons. 
The reason why we consider scenario B is that the relic density of the axion grows with heavy masses of the additional Higgs bosons (see Eq.~\eqref{eq:YaD scaling}). 

 In Fig~\ref{FIG:2HDMsC}, we show the excluded regions and future sensitivities of the various cosmological observations in the plane of $(m_a,g_{a\gamma\gamma})$ for the benchmark scenario A (top panels)  and B (bottom panels). 
 The results of Type-II (I) are shown in the left panels (right panels). 
 The constraints from X-ray, CMB spectral distortions, CMB anisotropy, and BBN observations are shown by the blue, gray, pink, and green-shaded regions, respectively. 
 Black solid contours show the fraction of the relic density for axion {${\cal R}_a$} (The definition is in Eq.~\eqref{eq:Ra definiton}) 
 and the dashed contours show the lifetime of the axion in the unit of the second. 
{As already discussed in Sec.~\ref{sec:delta_Neff}, our computation of axion productions from heavy Higgs does not contain the backreactions. 
To check the validity range of our computations for the ratio ${\cal R}_a$, we estimate the present number density of axion in the keV to sub-GeV scale when it is in thermal equilibrium: 
$n_{a,0}/S_0=45\zeta(3)/(2\pi^4g_{\ast S}(M_{H_2^{\prime}}))$.
this yields 
\bal \label{eq:rho_eq}
{\cal R}_a\simeq m_a\frac{45\zeta(3)}{2\pi^4}\frac{1}{g_{\ast S}(M_{H_2^{\prime}})}\frac{S_0}{\rho_{c0}\Omega_{\rm CDM}}\simeq {6.0}\left(\frac{m_a}{1{\rm keV}}\right)\;,
\eal
where in the last equality, we have used the present entropy $S_0\simeq{2.89}\times 10^3 {\rm cm}^{-3}$~{\cite{ParticleDataGroup:2024cfk}}, the present critical density $\rho_{c0}\simeq1.05\times 10^{-5} h^2 {\rm GeV} {\rm cm}^{-3}$~{\cite{ParticleDataGroup:2024cfk}} and the CDM density parameter $\Omega_{\rm CDM}=0.119h^{-2}$~{\cite{Planck:2018vyg}}.
The dotted regions show the region with ${\cal R}_a\gtrsim 6.2({m_a}/{1{\rm keV}})$. 
In this parameter, one should take into account the effect of the backreaction. 
In the case of Type-II, such a bound does not exist in {the region of} $g_{a\gamma}<1\times 10^{-9}{\rm GeV^{-1}}$ for benchmark scenarios A and B. 
}

X-ray observations are sensitive in the region of $m_a\lesssim 1{\rm MeV}$ since otherwise the decay channel $a\to e^+e^-$ opens and the decay rate $\Gamma_{a\to\gamma\gamma}$ is highly suppressed.  
Instead of the X-ray observations, the axion decays into an electron, and position contributes to the CMB anisotropies, which constrains the mass range of $m_a\gtrsim 1{\rm MeV}$. 
The difference between Type-II and Type-I for the excluded regions of X-ray and CMB anisotropy can be seen in the slope of the {upper} bound of $g_{a\gamma\gamma}$. 
For Type-II, with the scaling law concerning $m_a$ for decay rates and the fraction of the axion energy density, i.e., $\Gamma_{a\to\gamma\gamma}{\propto} m_a^3$, $\Gamma_{a\to ee}$,${\cal R}_a {\propto} m_a$, one can see that the photon flux increases with larger $m_a$. 
This means that the {upper} limit of $g_{a\gamma \gamma}$ tends to decrease for larger $m_a$. 
On the other hand, the {upper} limit of $g_{a\gamma \gamma}$ arising from the X-rays is almost independent of $m_a$ for Type-I. 
This characteristic behavior can be understood from the fact that the right-handed side of the Eq.~\eqref{eq:ineqxray} does not depend on $m_a$, with ${\cal R}_a\propto \frac{m_a}{v_a^2}\sim {\frac{1}{m_a^3}}g_{a\gamma\gamma}^2$  and ${\tau_a}\propto m_a^{-3}g_{a\gamma\gamma}^{-2}$ in case of $ m_a\lesssim 2m_e$. 
Furthermore, in contrast to Type-II, the {upper} bound of $g_{a\gamma\gamma}$ from CMB anisotropies moderates by $m_a$ for Type-I in the mass range of $m_a\gtrsim 2m_e$. 
This difference arises from the $m_a$ dependence on $g_{a\gamma\gamma}$ for a given total decay rate of $a$,$\Gamma_a$, 
\bal
\left(g_{a\gamma\gamma}^{\rm TyII}\right)^2=\frac{128 \alpha_{\rm em}^2\Gamma_a}{s_\beta^4\pi m_a^2 m_e^2 (1-4m_e^2/m_a^2)^{1/2}}\;, \quad
\left(g_{a\gamma\gamma}^{\rm TyI}\right)^2=\frac{m_a^3 \alpha_{\rm em}^2\Gamma_a}{18\pi m_e^6 (1-4m_e^2/m_a^2)^{1/2}}\;.
\eal
where we have assumed that $m_a\gtrsim2m_e$. 
The type dependence for the CMB spectral distortion and BBN bounds can also be understood from the above equation. 

 As can be understood from the analytical formulae, Eq.~\eqref{eq:ineqxray}, the 2HDM model parameter dependence for the X-ray bounds mostly stems from the fraction for the relic density {${\cal R}_a$}. 
 Since the decay rate $\Gamma_{H_2^\prime \to V a}$ is proportional to $s_{2\beta}^2$, the excluded region by the X-ray and CMB anisotropy observations tends to shrink by the large $t_{\beta}$.  
 On the contrary, $\Gamma_{H_2^\prime \to V a}$ increase by large $\tilde{m}_{12}$, so that the excluded regions are enraged by the large $\tilde{m}_{12}$. 
 Using this constraint from X-ray and CMB anisotropies for the decaying axion produced from the heavy Higgs decays, one may give the bound on the mass of the heavy Higgs bosons. 
 Let us assume that the axion of Type-II with $m_a$=1keV, $g_{a\gamma\gamma}\simeq 2\times 10^{-14}$ exists and the mixing angles for the Higgs bosons are fixed as $\tan{\beta}=5$ and $s_{\beta-\alpha}=1$. 
 If $\tilde{m}_{12} \gtrsim 30 $ TeV, such parameter region is excluded by the current X-ray bound. 
 Hence, to satisfy this constraint, the scale for the additional Higgs bosons $\tilde{m}_{12}$ should be smaller than $\tilde{m}_{12} \lesssim 30$ TeV.  

 One may consider relaxing X-ray and CMB bounds with a huge value of $t_\beta$, noting that the axion relic density is suppressed by $t_\beta$. 
 However, this does not work for Type-II since there is the upper bound of $t_\beta$ from $A\to\tau\tau$ (e.g., see Fig.10 of Ref.~\cite{Aiko:2020ksl}). 
 Even if additional Higgs bosons are too heavy to evade this bound at the LHC, the validity of the theory does not hold with high $t_\beta$ in taking into account the Landau pole constraint for the bottom Yukawa coupling. 
 We have checked that this happens in the case of $t_{\beta}\gtrsim 54$ at ${v_a}=10^{12}{\rm GeV}$ for Type-II.   

Future sensitivities of the X-ray and CMB anisotropies are shown by solid lines in Fig.~\ref{FIG:2HDMsC}. 
For X-ray observations, we here display the explored regions of  Athena~\cite{Thorpe-Morgan:2020rwc} (blue lines) and Theseus~\cite{Thorpe-Morgan:2020rwc} (Cyan and brown lines). 
For CMB anisotropies, those of $a\to e^+e^-$ and $a\to \gamma\gamma$ in Saimons array (SA)~\cite{Cang:2020exa} are shown by pink and purple lines. 
We here emphasize that other future X-ray observations such as XRISM~\cite{Zhou:2024sto} and eROSITA~\cite{Dekker:2021bos} can also give similar bounds for $g_{a\gamma\gamma}$. 
As seen from the figure, the sensitivities of $g_{a\gamma\gamma}$ can be better around one order magnitude regardless of $m_a$. 
The corresponding value of the fraction {${\cal R}_a$} for the axion energy density is the order of $ 10^{-15} (10^{-13})$ for Type-II (I) for $(m_a,m_H)=(10^4{\rm keV},600{\rm GeV})$, assuming the sensitivity of the SA of $a\to e^+e^-$. 
In this way, even if the axion abundance is considerably small, one can explore the axion thermally produced from the heavy Higgs bosons. 

{Before we move to conclusions, let us discuss the bounds from the cosmological observations and the direct detection for the axion-electron coupling $g_{ae}$. 
The cosmological constraints on $g_{ae}$ can be simply recast from the results of the axion-photon coupling $g_{a\gamma\gamma}$ discussed above through the relation $g_{ae}=-\frac{\pi X_{\phi e}m_e}{\alpha_{\rm EM}(E+E')}g_{a\gamma\gamma}$. 
For the constraint from the direct detection, we first introduce the ratio of the energy density of the axion and that of the observed dark matter,
\bal
\label{eq:def R_a 0}
{\cal R}_{a,0}\equiv\frac{\rho_{a,0}}{\rho_{\rm DM,0}}=e^{-t_U/\tau_a}{\cal R}_{a}. 
\eal
With this quantity, we calculate the upper bound of $g_{ae}$ by the inequality $\sqrt{ {\cal R}_{a,0} g_{ae}}<g_{ae}^{\rm exp.}$, where $g_{ae}^{\rm exp.}$ denotes the upper bound obtained by the direct detection of the axion in the dark matter direct searches such as XENONnT~\cite{XENON:2022ltv}. 
Let us compare the bounds on $g_{ae}$ evaluated from the cosmological observations and those of the direct detection.  
For Type-II, the upper bound of $g_{ae}$ obtained from the cosmological observations is stronger than the one from the direct detection. 
For Type-I, due to the absence of the anomalous contribution, the cosmological bound on $g_{ae}$ is moderate, and thus, the direct detection is compatible with the cosmological bounds in some parameter regions. 
These tendencies are illustrated in Fig.~\ref{FIG:2HDMsC_gaee} for scenario A and scenario B, where the black lines denote ${\cal R}_{a,0}$ and the dark gray lines show the lifetime of $a$.
{In the dotted region, the ${\cal R}_a$ does not satisfy Eq.~\eqref{eq:rho_eq}.}
The colored regions (lines) show the excluded (explored) regions from the current (future) cosmological observations and the current (future) direct detection experiment by the XENONnT~\cite{XENON:2022ltv} (LZ~\cite{LZ:2021xov}). 
While the cosmological constraints are shown by the same color as Fig.~\ref{FIG:2HDMsC}, the XENONnT and LZ results are shown by cyan regions and green lines, respectively. 
For Type-I, we also show the regions where the axion is overproduced by the orange regions. 
As can be seen from the figure, for scenario A and scenario B, the bounds of the axion relic abundance are more severe than the direct detection. 
Hence, for the axion produced from the heavy Higgs bosons with a mass higher than 600 GeV, the cosmological constraints are more important than that of the direct detection. 
}

\section{Conclusions}\label{sec:conclusion}
We have studied the thermal production of axions in DFSZ-type axion models, considering two different global U(1) PQ charge assignments presented in Table~\ref{tab:xia}.
Type-II predicts a typical axion with anomalous gauge boson couplings.
Type-I predicts an anomaly-free axion, for which the axion-photon coupling is generated by the threshold effect of a fermion in the loop.
It should be noted that the thermal axion production discussed in this paper can be applied to the original QCD axion model, as the discussion does not rely on the axion mass.

For the thermal history of the universe, we have assumed that the reheating temperature ($T_R$) is in the range $m_H \lesssim T_R \lesssim v_a$. Under these conditions, the heavy Higgs bosons are present in the thermal bath of the early universe.
We have comprehensively calculated thermal axion production in the symmetric phase via the trilinear Higgs couplings $\lambda_{H'_1 H'_2 a}$ (see Eq.\eqref{eq:coupH1H2a}), which generates heavy Higgs boson decays $H'_2\to H'_1 a$ and scattering processes in Eq.\eqref{eq:scaterings}.
We have compared these processes with axion production from the SM-like Higgs boson decay and top-quark scatterings as representative axion production processes from SM particles in the broken phase.
We have found that $H'_2\to H'_1 a$ consistently gives dominant contributions to axion thermal production.

{
To illustrate how the axion affects the evolution of the early Universe, we have calculated its contributions to the effective number of neutrino species, $\Delta N_{\rm eff}$, focusing on axions with sufficiently small mass.
This discussion can be applied to canonical DFSZ QCD axion models.
} {We have found that the axion abundance could reach the equilibrium value at temperatures higher than the weak scale,  $\Delta N_{\rm eff} \simeq 0.03$, can be reached for the decay constant significantly higher than previously thought, as high as $v_a = {\cal O}(10^9)\,$GeV. }
We have also discussed the cosmological implications, focusing on the axion mass range of the keV to sub-GeV scale.
The relevant constraints come from X-ray observations, CMB anisotropy, SDs, and BBN. The combination of these limits provides a stringent upper bound on the axion-photon coupling, as illustrated in Fig.~\ref{FIG:2HDMsC}.
In particular, even if the fraction parameter ${\cal R}_a$ of the axion energy density is of the order ${\cal O}(10^{-10}\mbox{-}10^{-13})$, such parameter regions are still constrained by X-ray or CMB observations.

The axion production from heavy Higgs decays evaluated in this paper is based on the highly plausible assumption that the reheating temperature is higher than the heavy Higgs mass and the Higgs sector is thermalized. We emphasize here that the axion production from heavy Higgs decays gives the dominant contribution, which was overlooked in the literature on the thermal production of axions.
Consequently, the constraints on axions derived here from various observations can be considered extremely general.
In particular, the estimate of $N_{\rm eff}$ is applicable to the DFSZ QCD axion, and it
will enable us to derive {stringent constraints on the model parameters of the axion and Higgs sectors combined with the future observations.} 
These results provide a valuable cosmological constraint on DFSZ-type axions.

In conclusion, our comprehensive study of axion thermal production in DFSZ-type models, combined with the analysis of cosmological implications, offers robust and widely applicable constraints on axion parameters.

\acknowledgments
The present work is supported by JSPS Core-to-Core Program (grant number: JPJSCCA20200002) (F.T.), JSPS KAKENHI Grant Numbers 20H01894 (F.T.), 20H05851 (F.T.), and 21K20363 (K.S.). 
This article is based upon work from COST Action COSMIC WISPers CA21106, supported by COST (European Cooperation in Science and Technology).

\appendix
\section{Relation for the parameters for the Higgs potential}
\subsection{Mass parameters and quartic couplings in the Higgs basis }\label{ap:HBparameter}
The Higgs potential in the Higgs basis is given by Eq.~\eqref{eq:VHB}. 
The mass parameters $Y_{i}$ and the couplings $Z_{j}$ are presented in terms of the original potential parameters as
\begin{align}
  Y_{1}^2&=m_{11}^2c_{\beta}^2+m_{22}^2s_{\beta}^2-m_{12}^2s_{2\beta}\,,
  \label{maa}\\
  Y_{2}^2&=m_{11}^2s_{\beta}^2+m_{22}^2c_{\beta}^2+m_{12}^2s_{2\beta}\,,
  \label{mbb}\\
  Y_{3}^2 &=
  \frac{1}{2}(m_{11}^2-m_{22}^2)s_{2\beta}+m_{12}^2c_{2\beta}\,,\label{mab} \\
  \!\!\!\!\!\!\!\!\!\!\!\!\!\!Z_1&=
  2\lambda_1c_{\beta}^4+2\lambda_2s_{\beta}^4+\frac{1}{2}\lambda_{34}s_{2\beta}^2\,,
  \label{Lam1def}  \\
  \!\!\!\!\!\!\!\!\!\!\!\!\!\!Z_2 &=
  2\lambda_1s_{\beta}^4+2\lambda_2c_{\beta}^4+\frac{1}{2}\lambda_{34}s_{2\beta}^2
  \,,
  \label{Lam2def}      \\
  \!\!\!\!\!\!\!\!\!\!\!\!\!\!Z_3 &=
  \frac{1}{4}s_{2\beta}^2\left[2\lambda_1+2\lambda_2-2\lambda_{34}\right]
  +\lambda_3\,,
  \label{Lam3def}      \\
  \!\!\!\!\!\!\!\!\!\!\!\!\!\!Z_4 &=
  \frac{1}{4}s_{2\beta}^2\left[2\lambda_1+2\lambda_2-2\lambda_{34}\right]
  +\lambda_4\,,
  \label{Lam4def}      \\
  \!\!\!\!\!\!\!\!\!\!\!\! Z_5 &=
  \frac{1}{4}s_{2\beta}^2\left[2\lambda_1+2\lambda_2-2\lambda_{34}\right]\,,
  \label{Lam5def}      \\
  \!\!\!\!\!\!\!\!\!\!\!\!Z_6 &=
  -\frac{1}{2}s_{2\beta}\left[2\lambda_1c_{\beta}^2
  -2\lambda_2s_{\beta}^2-\lambda_{34}c_{2\beta}\right]\,,
  \label{Lam6def}      \\
  \!\!\!\!\!\!\!\!\!\!\!\!Z_7 &=
  -\frac{1}{2} s_{2\beta}\left[2\lambda_1s_{\beta}^2-2\lambda_2c_{\beta}^2+
  \lambda_{34} c_{2\beta}\right]\,,
  \label{Lam7def}
\end{align}
with $\lambda_{34}$ being $\lambda_{34}=\lambda_3+\lambda_4$. 

\subsection{Mass parameters and quartic couplings in the mass basis of the symmetric phase} \label{ap:mass coupling in MB SP}
We present the potential parameters in the mass basis in the symmetric phase with the model parameters Eq.~\eqref{eq:inputs} and the mixing angle $\omega$. 
They are given by
\bal
  Y^{\prime2}_1&=\frac{1}{4 s_\beta c_\beta c_{2\omega}}\Bigg[
    m_{h}^2 \left(c_{2\beta} s_{2(\beta-\alpha)} s^2_{\omega}-s_{2\beta} s^2_{\beta-\alpha} c_{2\omega}\right)
     \notag \\
    &+m_{H}^2 \left(-s_{2\beta} c^2_{\beta-\alpha} c_{2\omega}-c_{2\beta} s_{2(\beta-\alpha)} s^2_{\omega}\right)-4 \tilde{m}^2_{12} s_{\beta} c_{\beta} s^2_{\omega}
  \Bigg]\;, \\
  Y^{\prime2}_2&=\frac{1}{4 s_\beta c_\beta c_{2\omega}}\Bigg[
  m_{h}^2 \left(-c_{2\beta} s_{2(\beta-\alpha)} c^2_{\omega}-s_{2\beta} s^2_{\beta-\alpha} c_{2\omega}\right) \notag \\
  &+m_{H}^2 \left(c_{2\beta} s_{2(\beta-\alpha)} c^2_{\omega}-s_{2\beta} c^2_{\beta-\alpha} c_{2\omega}\right)
  +4 \tilde{m}^2_{12} s_{\beta} c_{\beta} c^2_{\omega}
    \Bigg]\;,
\eal
\bal
  Z^{\prime}_1&=\frac{1}{4v^2c_\beta^2s_\beta^2c_{2\omega}}\Bigg[
    m_{h}^2 c_{2\omega} (c_{\alpha+\beta}-c_{\alpha-\beta} c_{2(\beta+\omega)})^2 \notag \\
    &+m_{H}^2 c_{2\omega} (s_{\alpha+\beta}-s_{\alpha-\beta} c_{2(\beta+\omega)})^2-4 \tilde{m}^2_{12} s^2_{\omega} c_{2\omega} s^2_{2\beta+\omega}
  \Bigg]
  \;,\\
  Z^{\prime}_2&=\frac{1}{4v^2c_\beta^2s_\beta^2c_{2\omega}}\Bigg[
  m_{h}^2 c_{2\omega} (c_{\alpha-\beta} c_{2(\beta+\omega)}+c_{\alpha+\beta})^2 \notag \\
  &+m_{H}^2 c_{2\omega} (s_{\alpha-\beta} c_{2(\beta+\omega)}+s_{\alpha+\beta})^2
  -4 \tilde{m}^2_{12} c^2_{\omega} c_{2\omega} c^2_{2\beta+\omega}
  \Bigg]
  \;,\\
  Z^{\prime}_3&=\frac{1}{16v^2 s_\beta^2c_\beta^2}\Bigg[
  m_{h}^2 (-c_{2(\alpha-\beta)} \{c_{4(\beta+\omega)}+1\}+2 c_{2(\beta+\alpha)}-c_{4(\beta+\omega)}+1) \notag \\
  &+m_{H}^2 (c_{2(\alpha-\beta)} \{c_{4(\beta+\omega)}+1\}-2 c_{2(\beta+\alpha)}-c_{4(\beta+\omega)}+1) \notag \\
  &+2 \tilde{m}^2_{12} (c_{4(\beta+\omega)}+c_{4\beta}-2)
 \Bigg] \;,\\
  Z^{\prime}_4&=\frac{1}{16v^2 s_\beta^2c_\beta^2}\Bigg[
 4 m_{h}^2 c^2_{\alpha-\beta} s^2_{2(\beta+\omega)}
+4 m_{H}^2 s^2_{\alpha-\beta} s^2_{2(\beta+\omega)} \notag \\
&+\tilde{m}^2_{12} (2 c_{4(\beta+\omega)}-4 c_{4\beta}+2)  
 \Bigg] \;,\\
  Z^{\prime}_5&=\frac{1}{16v^2 s_\beta^2c_\beta^2}\Bigg[
  4 m_{h}^2 c^2_{\alpha-\beta} s^2_{2(\beta+\omega)}
  +4 m_{H}^2 s^2_{\alpha-\beta} s^2_{2(\beta+\omega)} 
  -4 \tilde{m}^2_{12} s^2_{2(\beta+\omega)} \Bigg]
  \;,
\eal
\bal
  Z^{\prime}_6&=\frac{1}{16v^2 s_\beta^2c_\beta^2}\Bigg[
  4 m_{h}^2 c_{\alpha-\beta} s_{2(\beta+\omega)} (c_{\alpha+\beta}-c_{\alpha-\beta} c_{2(\beta+\omega)}) \notag \\
  &+4 m_{H}^2 s_{\alpha-\beta} s_{2(\beta+\omega)} (s_{\alpha+\beta}-s_{\alpha-\beta} c_{2(\beta+\omega)})
  +4 \tilde{m}^2_{12} s_{\omega} (c_{4\beta+3\omega}-c_{\omega})
  \Bigg]\;,\\
  Z^{\prime}_7&= \frac{1}{16v^2 s_\beta^2c_\beta^2}\Bigg[
  4 m_{h}^2 c_{\alpha-\beta} s_{2(\beta+\omega)} (c_{\alpha-\beta} c_{2(\beta+\omega)}+c_{\alpha+\beta}) \notag \\
  &+4 m_{H}^2 s_{\alpha-\beta} s_{2(\beta+\omega)} (s_{\alpha-\beta} c_{2(\beta+\omega)}+s_{\alpha+\beta})
  -4 \tilde{m}^2_{12} c_{\omega} (s_{4\beta+3\omega}+s_{\omega})
  \Bigg]\;.
\eal

\section{Thermal corrected masses} \label{ap:thermal mass}
In the symmetric phase, thermal corrections to particle masses are not negligible. 
We here describe the thermal corrected masses of Higgs bosons in the Higgs basis $(H_1,H_2)$. 
The mass matrix ${\cal M}_{H_1H_2}$ with thermal corrections is written by 
\bal \label{eq:mass_matrix_w_thermal_corrections}
{\cal M}_{H_1 H_2,T}
=
\begin{pmatrix}
  Y_1^2 && -Y_3^2\\  
 -Y_3^2 && Y_2^2
\end{pmatrix}
+
\begin{pmatrix}
  \Pi_{H_1 H_1} && \Pi_{H_1 H_2}\\  
  \Pi_{H_1 H_2} && \Pi_{H_2 H_2}
\end{pmatrix} \;,
\eal
where each component of the thermal corrections is given by
\bal
\Pi_{H_1 H_1}&=T^2
\left[\frac{1}{4}Z_1+\frac{1}{6}Z_3+\frac{1}{12}Z_4+\frac{3}{16}g^2+\frac{1}{16}g^{\prime 2}+\frac{1}{4}y_{t,1}^2+\frac{1}{4}y_{b,1}^2+\frac{1}{12}y_{\tau,1}^2\right]\;, \\ 
\Pi_{H_2 H_2}&=T^2
\left[\frac{1}{4}Z_2+\frac{1}{6}Z_3+\frac{1}{12}Z_4+\frac{3}{16}g^2+\frac{1}{16}g^{\prime 2}+\frac{1}{4}y_{t,2}^2+\frac{1}{4}y_{b,2}^2+\frac{1}{12}y_{\tau,2}^2\right]\;, \\ 
\Pi_{H_1 H_2}&=T^2
\left[\frac{1}{4}Z_6+\frac{1}{4}Z_7+\frac{1}{4}y_{t,1} y_{t,2}+\frac{1}{4}y_{b,1} y_{b,2}+\frac{1}{12}y_{\tau,1}y_{\tau,2}\right]\;,  
 \eal
 where $y_{f,1}$ $(f=t,b,\tau)$  denotes the Yukawa coupling for $H_{1}$ and 
 the coupling $y_{f,2}$ does the one for $H_{2}$. 
 They are given by $y_{f,1}=\sqrt{2}m_f/v$ and $y_{f,2}=\sqrt{2}\zeta_f m_f/v$. 
 The factor $\zeta_f$ depends on the type of Yukawa interactions: 
 $\zeta_t=\zeta_b=\zeta_\tau=\cot\beta$ for Type-I; $\zeta_t=\cot\beta, \zeta_b=\zeta_\tau=-\tan\beta$ for Type-II.  
 
 We calculate the thermal corrected masses $M_{H'_1}^2$ and $M_{H'_2}^2$ by diagonalizing the thermal corrected mass matrix ${\cal M}_{H_1H_2, T}$. 
 On the other hand, we omit the effect of the thermal mass corrections for the mixing of the Higgs bosons. 
 Thus, we use the mixing angle at the zero temperature, $\omega$. 
 Indeed, the effect of thermal corrections for the mixing angle is not large for the parameter region that we are interested in. 
 we have numerically evaluated the thermal corrected mixing angles ($\omega_T$) from Eq.~\eqref{eq:mass_matrix_w_thermal_corrections} and checked that the difference $\omega_T-\omega$ is $0.03$ at $(T,s_{\beta-\alpha},t_\beta)=(m_H,1,5)$ for any value of ${\tilde{m}_{12}}$ in the degenerate mass scenario $m_H=m_{H^\pm}={\tilde{m}_{12}}$.

\bibliographystyle{JHEP}
\bibliography{references}
\end{document}